\documentclass[pre,eqsecnum,floatfix,showpacs,amsmath,amssymb,longbibliography,superscriptaddress,aps,prf]{revtex4-2}
\usepackage{graphicx}
\usepackage{ulem}
\usepackage{bm}
\usepackage{xcolor}
\usepackage{float}
\usepackage{caption}
\usepackage{subcaption}
\usepackage{mymacros}

\begin{document}

\title{Retraction Dynamics of a Highly Viscous Liquid Sheet}

\author{Taosif Ahsan}
\thanks{T.A.\ and R.B.\ contributed equally to this work.}
\affiliation{Department of Physics, Massachusetts Institute of Technology, Cambridge, MA 02139, USA}
\affiliation{Department of Physics, Princeton University, Princeton, NJ 08544, USA}

\author{Rodolfo Brand\~ao}
\thanks{T.A.\ and R.B.\ contributed equally to this work.}
\affiliation{School of Mathematics, University of Bristol, Fry Building, Woodland Road, Bristol BS8 1UG, UK}
\affiliation{Department of Mechanical and Aerospace Engineering, Princeton University, Princeton, NJ 08544, USA}

\author{Benny Davidovitch}
\affiliation{Department of Physics, University of Massachusetts Amherst, Amherst, Massachusetts 01003, USA}

\author{Howard A. Stone}
\affiliation{Department of Mechanical and Aerospace Engineering, Princeton University, Princeton, NJ 08544, USA}

\begin{abstract}

We study the one-dimensional capillary-driven retraction of a finite planar liquid sheet in the asymptotic regime where both the initial length-to-thickness ratio $l_0/h_0$ and a characteristic Ohnesorge number, $\text{Oh}$, which is inversely related to the Reynolds number of the flow, are large. In this regime, the fluid domain decomposes into two regions: a thin-film region governed by one-dimensional mass and momentum equations, and a small tip region near the free edge described by a self-similar Stokes flow. Asymptotic matching between these regions yields an effective boundary condition for the thin-film region, representing a balance between viscous and capillary forces at the free edge. We find that surface tension drives the thin-film flow only through this boundary condition, while the local momentum balance is dominated by viscous and inertial stresses. We show that the thin-film flow possesses a conserved quantity, reducing the governing equations to a one-dimensional free-boundary problem in which the sheet thickness satisfies the heat equation with time-dependent boundary conditions. The reduced problem depends on a single dimensionless parameter $\mathcal{L} = l_0 / (4 h_0 \text{Oh})$. Numerical solutions of the reduced model agree well with previous studies based on different formulations, including full Navier–Stokes simulations, and reveal that the sheet undergoes distinct retraction regimes depending on $\mathcal{L}$ and a dimensionless time after rupture $T$. We derive asymptotic approximations for the thickness profile, velocity profile, and retraction speed during the early and late stages of retraction. At early times, the retraction speed grows as $T^{1/2}$, characteristic of diffusive processes, while at late times it decays as $1/T^2$ due to the finiteness of the sheet. Both results are consistent with previous findings. We also analyze an intermediate regime, not previously studied in detail, which arises for very long sheets ($\mathcal{L} \gg 1$). During this phase, the retraction speed approaches the classical Taylor–Culick value, despite the sheet not exhibiting a large circular rim as in classical studies of film rupture. When $T \approx \mathcal{L}$, the speed undergoes a sudden deceleration, which we characterize numerically, going from the Taylor–Culick speed to that predicted by the late-time asymptotics.

\end{abstract}

\maketitle


\section{Introduction}

After a liquid sheet ruptures, its free edge retracts under the action of capillary forces. This phenomenon is commonly seen in bursting bubbles and is central to many natural and industrial processes, such as fluid atomization \cite{bremond:07,Villermaux:20}, spread of pathogens in water \cite{Bourouiba:21}, and curtain coating applications \cite{finnicum:93}.

The retraction speed of a ruptured liquid sheet was independently derived by Taylor \cite{Taylor:59} and Culick \cite{Culick:60}. They considered an idealized scenario of an infinitely extended, two-dimensional planar sheet composed of an inviscid liquid. By assuming that the liquid accumulates in a circular rim at the free edge and applying conservation of mass and momentum at the rim, they found a constant retraction speed given by
\begin{equation}
u_{\text{TC}} = \sqrt{\frac{\gamma}{\rho h_0}},
\label{eq: taylor culick speed}
\end{equation}
where $\gamma$ is the surface tension, $\rho$ is the liquid density, and $2h_0$ is the uniform film thickness prior to rupture. The predicted speed---now often referred to as the `Taylor–Culick speed'---shows good agreement with experiments with low-viscosity liquids such as water \cite{Ranz:59, McEntee:69}.

In contrast, later experiments by Debr\'egeas and coworkers \cite{Debregeas:95, Debregeas:98} on the growth of circular holes in highly viscous sheets revealed a markedly different dynamics \cite{Debregeas:95, Debregeas:98}. Notably, the liquid did not accumulate at the rim; instead, the film thickness remained spatially uniform during retraction. Furthermore, the retraction speed was not constant as in \eqref{eq: taylor culick speed} but increased over time on a characteristic viscous timescale proportional to $\mu h_0 / \gamma$, where $\mu$ is the liquid viscosity. 

Motivated by these experiments, several theoretical works examined the retraction dynamics of highly viscous sheets \cite{Brenner:99, Song:99, Sunderhauf:02, Savva:09, Deka:20}. An influential numerical study was reported  by Brenner and Gueyffier \cite{Brenner:99}, who employed a thin-film model to analyze the retraction of a two-dimensional planar sheet. In their model, the flow was governed by two coupled partial differential equations for the mass and momentum balances in the liquid, with the latter including inertial, viscous, and capillary stresses. The dynamics depended on two dimensionless groups,
\begin{equation}
\frac{l_0}{h_0}, \qquad \text{Oh}= \frac{\mu}{\sqrt{\rho h_0 \gamma}},
\label{eq: parameters}
\end{equation}
where $l_0$ is the initial length of the sheet. The first parameter is the initial aspect ratio of the sheet, assumed large, while the second is the Ohnesorge number, which is inversely related to the Reynolds number of the flow.

We note that since the simulations in \cite{Brenner:99} considered planar sheets, they could not capture the precise dynamics seen in expanding holes.  Specifically, at large $\text{Oh}$, corresponding to highly viscous sheets, the dynamics does not exhibit the exponential retraction speed that was observed in experiments \cite{Debregeas:95}; instead, it depends on the relative magnitude of $\text{Oh}$ and $l_0/h_0$. When $\text{Oh} \gg l_0/h_0$, the sheet thickness remains spatially uniform throughout the retraction, in a manner resembling the experiments in \cite{Debregeas:95,Debregeas:98}, but with a qualitatively different speed. As $\text{Oh}$ decreases relative to $l_0/h_0$, liquid accumulation in the rim becomes more pronounced, and, for $1 \ll \text{Oh} \ll l_0/h_0$, they observed the formation of a circular rim moving at the Taylor–Culick speed, as in the nearly inviscid case.

Building on the work of Brenner and Gueyffier \cite{Brenner:99}, later studies adopted similar thin-film models to investigate the retraction of viscous sheets in various settings \cite{Sunderhauf:02, Savva:09, Gordillo:11, Deka:20}. In particular, Savva and Bush \cite{Savva:09} simulated the retraction of flat sheets with both planar and circular geometries in the limit of infinitely long films ($l_0/h_0 = \infty$). Consistent with \cite{Brenner:99}, they found that in both geometries and for all values of $\text{Oh}$, the free edge eventually reached the Taylor–Culick speed, although the approach was slower for larger $\text{Oh}$. Their numerical results were complemented by an early-time analysis of the velocity profile. To this end, they assumed that the sheet initially consisted of a small semi-circular cap at the tip patched to a flat film. For the circular geometry---\textit{i.e.}, a circular hole expanding in a film---they found that the hole radius initially increased exponentially, in agreement with the theory and experiments in \cite{Debregeas:95}.

More recently, Deka and Pierson \cite{Deka:20} examined the regime where sheet is long but finite and highly viscous. Starting from the thin-film equations of \cite{Brenner:99},
and making a series of approximations in the regime $1\ll l_0/h_0 \ll \text{Oh}$, they derived a 
similarity solution valid far from the tip and at late times after rupture. Their solution predicts that the sheet remains spatially uniform, with its thickness growing linearly in time. By enforcing mass conservation and making a few assumptions on the sheet geometry near the tip, they also derived an analytic approximation for the retraction speed. To support their analysis, the authors conducted full Navier–Stokes simulations for a range of large $l_0/h_0$ and $\text{Oh}$, finding good agreement with their similarity solution as well as earlier simulations and theories.

Although these studies show that thin-film models offer a powerful framework for understanding liquid-sheet retraction, all of the above-mentioned models share an undesirable feature. As Brenner and Gueyffier \cite{Brenner:99} and others have noted, the core assumption of thin-film theory, namely that interface slopes are small, inevitably breaks down near the highly curved free edge. Furthermore, to avoid singularities at the edge, the capillary stresses in the momentum equation must be regularized by including the full nonlinear curvature.

\begin{figure}[t!]
\includegraphics[width=0.5\textwidth]{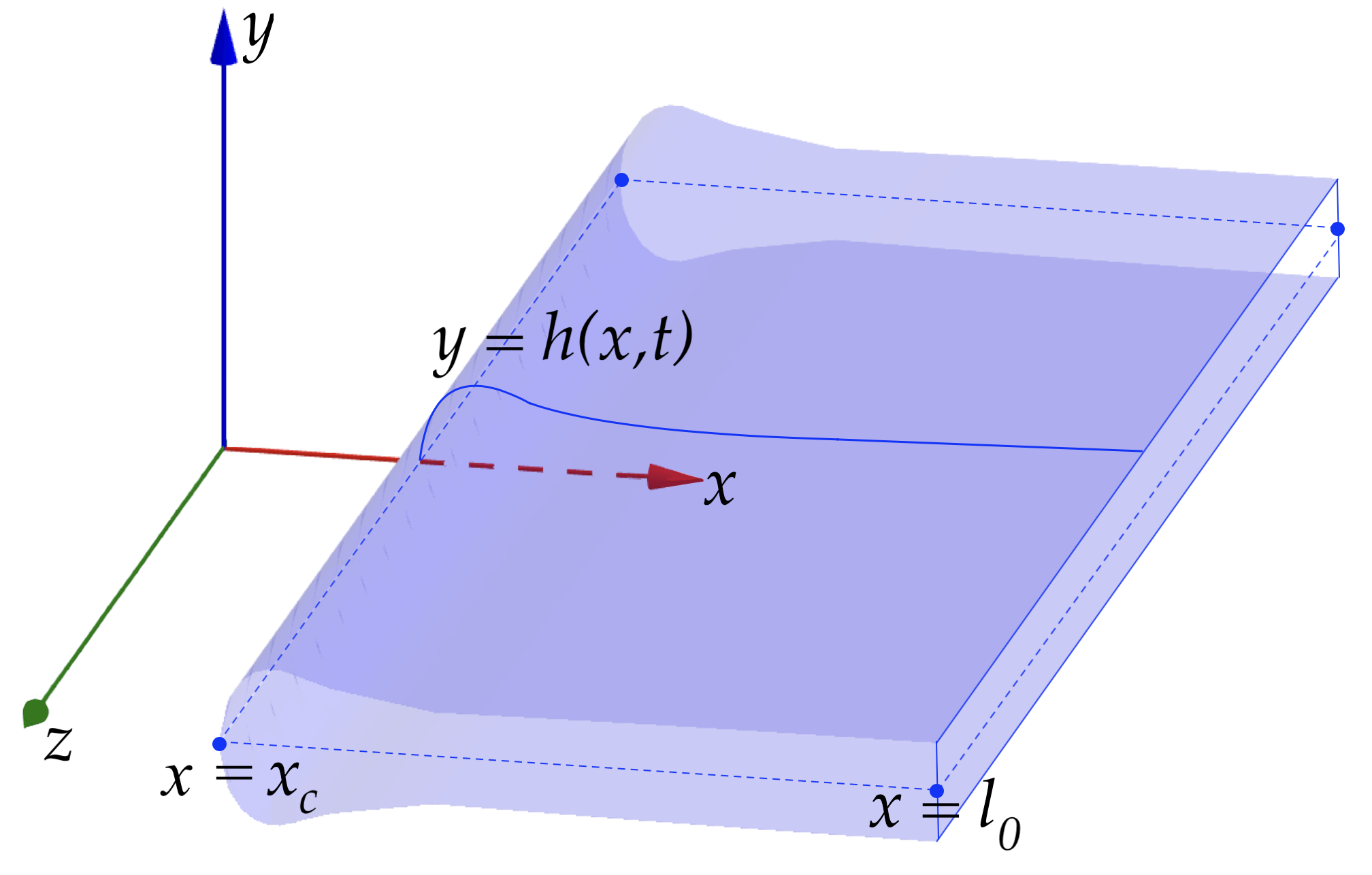}
\caption{Schematic of a ruptured planar thin sheet. A flat surface across the $x$–$z$ plane is fixed at $x = l_0$ and ruptured along the line $x = 0$. As time $t$ progresses, the free edge moves to $x = x_c(t)$. The half-thickness along the $y$-axis is given by $h(x, t)$ and shown as a continuous blue curve.}
\label{fig: Thinfilm}
\end{figure} 

This regularization not only raises concerns about the validity of thin-film models near the tip but also increases the complexity of the governing equations, thereby hindering both numerical and asymptotic analyses. For example, to avoid directly addressing the highly curved tip region, both Savva and Bush \cite{Savva:09} and Deka and Pierson \cite{Deka:20} made simplifying assumptions in their respective early-time and self-similar analyses, some of which were not fully justified.

In this paper, we revisit the problem of the retraction of a two-dimensional planar liquid sheet, focusing on the limit of highly viscous ($\text{Oh} \gg 1$) and long ($l_0/h_0 \gg 1$) sheets. We demonstrate that the fluid domain decomposes into two distinct asymptotic regions: a large sheet-scale region, governed by the thin-film approximation, and a small region near the free edge, governed by the two-dimensional Stokes equations.

Within the thin-film region, we show that capillary stresses are negligible relative to viscous and inertial forces and enter the problem solely through a boundary condition at the free edge.  This greatly simplifies the equations, which now depend only on a single dimensionless parameter proportional to the ratio of $l_0/h_0$ and $\text{Oh}$, and removes the need for any regularization. We note that our asymptotic description is closely related to that of Munro \textit{et al.} \cite{Munro:15} in their study of bubble coalescence, although the geometry of the liquid film in their work differs from that considered here.

We identify a conserved quantity within the thin-film equations, which leads to a linear relationship between the liquid velocity and the interface slope. This reduces the original coupled equations for mass and momentum balances to a single equation, which may be written solely in terms of either the thickness or velocity. In terms of thickness, the reduced problem takes the form of a one-dimensional heat equation with time-dependent boundary conditions. Alternatively, in terms of velocity, it reduces to Burgers' equation, also with time-dependent boundary conditions.

We perform numerical simulations of the reduced problem for the thickness, verifying that it accurately captures the distinct retraction behaviors observed in prior numerical studies \cite{Brenner:99, Sunderhauf:02, Savva:09, Deka:20}. Additionally, we systematically analyze various limiting cases of our model, recovering previously derived asymptotic solutions and obtaining new analytical approximations in regimes that have not been previously considered.

\section{Problem statement}
\label{sec: statement}
\subsection{Thin-film model}
\label{sec: thin film model}

We describe the geometry of the liquid sheet using the Cartesian coordinate system $(x,y,z)$ shown in Fig.~\ref{fig: Thinfilm}. The sheet is assumed to be symmetric about the $xz$ plane and invariant in the $z$-direction, thus rendering the problem effectively two dimensional. Accordingly, we henceforth restrict our attention to the $x-y$ plane. The sheet is initially in equilibrium with length $l_0$ and uniform thickness $2 h_0$. When it ruptures at one end, say located at $x=0$, capillary forces cause it to retract towards the opposite end, say fixed at $x=l_0$. Denoting the time after rupture by $t$, the liquid domain is
\begin{equation}
-h(x,t) < y < h(x,t),\qquad x_c(t) < x < l_0,
\end{equation}
where $h(x,t)$ is the thickness profile and $x_c(t)$ is the $x-$coordinate of the free edge, which satisfy the initial conditions 
\refstepcounter{equation}
$$
\label{eq: init cond dim}
h(x,0)=h_0,\qquad x_c(0)=0.
\eqno{(\theequation\mathrm{a},\mathrm{b})}.
$$

Assuming, subject to {\it a posteriori} verification, that $|\partial h/\partial x|\ll1$, the liquid flow can be described using the standard thin-film approximation \cite{Erneux:93,Howell:96}. Accordingly, following the Trouton approximation \cite{Trouton1906}, we
note that to leading order in $h_0/l_0$ the fluid velocity is in the $x$-direction, 
and depends exclusively on $x$ and $t$, namely,  ${\bf v} \approx v(x,t) \hat{e}_x$. As the sheet is initially at equilibrium, we impose the initial condition
\begin{equation}
v(x,0)=0.
\label{eq: init vel}
\end{equation}
The thickness and velocity profiles are governed by the continuity equation
\begin{equation}
\frac{\partial h}{\partial t} + \frac{\partial}{\partial x}\left(v h\right)=0
\label{eq: continuity}
\end{equation}
and the momentum equation
\begin{equation}\label{eq: dimen mom}
\underbrace{\rho h \frac{D v}{Dt}}_{\text{inertial}} =  
\underbrace{4 \mu\frac{\partial}{\partial x}\left(h \frac{\partial v}{\partial x}\right)}_{\text{viscous}} + 
\underbrace{\gamma h \frac{\partial^3 h}{\partial x^3}}_{\text{capillary}},
\end{equation}
wherein $D/Dt = \partial/\partial t + v \partial/\partial x$ is the material derivative. 
Eqs.\,\eqref{eq: continuity} and  \eqref{eq: dimen mom} provide a description of a thin liquid sheet that includes inertia, surface tension, and viscous stresses. 
However, as we shall see in Sec.\,\ref{sec: scales}, scaling arguments reveal that capillary stresses can be systematically neglected in the large-Ohnesorge regime.

It remains to prescribe boundary conditions for the problem. At the free edge, the velocity $v$ and retraction speed $dx_c/dt$ are related by the kinematic boundary condition,
\begin{equation}
\frac{d x_c}{dt}=v\left(x_c(t),t\right).
\label{eq: dimens kine}
\end{equation}
One might intuitively anticipate an additional boundary condition prescribing the thickness profile at the free edge, as done in \cite{Brenner:99}. There, it was assumed that the sheet has a circular tip where $h\propto \left(x_c - x\right)^{1/2}$ as $x\to x_c$. Such a condition, however, is inconsistent with the thin-film approximation, as it implies that the slope at $x=x_c$ behaves as $\partial h/\partial x\propto1/\left(x_c - x\right)^{1/2}$, and thus diverges at $x=x_c$. To systematically derive a second boundary condition, we instead decompose the liquid domain into two asymptotically distinct regions. The first is a long `thin-film' region, where $|\partial h/\partial x| \ll 1$ and Eqs.\,\eqref{eq: continuity} and \eqref{eq: dimen mom} apply. The second is a small tip region near the free edge, with a characteristic lengthscale of $\mathcal{O}(h)$, where the full two-dimensional Stokes flow must be calculated. With this distinction, the point $x = x_c(t)$ is interpreted as the approximate position of the free edge on the thin-film scale. In Appendix \ref{app: tip region}, we use matched asymptotic expansions \cite{Hinch:book} to derive an \textit{effective} boundary condition for the thin-film region, which reads
\begin{equation}
4\mu h\left(x_c(t), t\right)\frac{\partial v}{\partial x}\left(x_c(t), t\right)=
-\gamma.
\label{eq: dimens dyn}
\end{equation}
The above condition can also be rationalized by a force balance argument, as done in \cite{Munro:15}. The left- and right-hand sides of Eq.\,\eqref{eq: dimens dyn} represent, respectively, the viscous and capillary forces (per unit length) acting on the tip region. These must balance, as inertia is negligible in that region provided the tip remains sufficiently small relative to the length of the film. This description, in turn, requires a large Ohnesorge number ($\text{Oh} \gg 1$) and that the elapsed time is not asymptotically large relative to a characteristic viscous timescale, as discussed in detail in Appendix~\ref{app: tip region}.

At the fixed end, we impose the impermeability condition
\begin{equation}
v(l_0,t) = 0 
\label{eq: no flow cond}
\end{equation}
and the zero-slope, or symmetry, condition
\begin{equation}
\frac{\partial h}{\partial x}(l_0, t)  = 0 .
\label{eq: no slope cond}
\end{equation}
According to \cite{Brenner:99,Deka:20}, these conditions mimic those during the bursting of an axisymmetric bubble, in which case $x=l_0$ lies in the symmetry axis of the bubble.

\subsection{Scaling analysis}
\label{sec: scales}

We now perform a scaling analysis of the thin-film region to estimate the magnitude of the stresses in Eq.\,\eqref{eq: dimen mom}. We denote the characteristic scales for the sheet thickness and velocity, and for their variations in the horizontal direction, and in time by $h^*$, $v^*$, $l^*$, and $t^*$, respectively. 

The flow is driven by the dynamic boundary condition \eqref{eq: dimens dyn}, which implies
\begin{equation}
\frac{4\mu h^* v^*}{l^*} = \gamma.
\label{eq: scale bc}
\end{equation}
From this, the viscous term in Eq.\,\eqref{eq: dimen mom} scales as $\frac{4\mu h^* v^*}{l^{*2}} = \frac{\gamma}{l^*}$, while the capillary term scales as $\gamma \frac{h^{*2}}{l^{*3}}$, such that
\begin{equation}\label{eq: capil negl}
\frac{\text{capillary stresses}}{\text{viscous stresses}} = \left(\frac{h^*}{l^*}\right)^2 \ll 1.
\end{equation}
Thus, viscous stresses dominate over capillary stresses in the thin-film region. The same conclusion was reached by Munro \textit{et al.} \cite{Munro:15} in their analysis of the thin-sheet flow between coalescing bubbles. Physically, this means that while the flow is driven by surface tension acting on the highly curved tip,  capillary stresses associated with thickness variations on the thin-film scale---\textit{i.e.} far away from the tip---are negligible compared to viscous stresses. 
We emphasize that this argument depends on Eq.\,\eqref{eq: dimens dyn}, which requires $\text{Oh} \gg 1$.

We proceed by discussing the role of inertia in the problem. Despite the sheet being highly viscous, inertia is dominant in Eq.\,\eqref{eq: dimen mom} during many stages of the retraction.  To see this, note that in the absence of inertia, Eq.\,\eqref{eq: dimen mom} can be integrated in conjunction with Eq.\,\eqref{eq: dimens dyn},  leading to a uniform viscous stress across the film, $4 \mu h \frac{\partial v}{\partial x} = -\gamma$. However, this result is incompatible with the initial conditions Eq.\,\eqref{eq: init cond dim} and \eqref{eq: init vel}. For infinitely long sheets $(l_0=\infty)$, as studied by Savva and Bush \cite{Savva:09}, the purely viscous solution also fails to satisfy the boundary conditions of a flat film and vanishing flow at infinity, given by Eqs.\,\eqref{eq: no flow cond} and \eqref{eq: no slope cond}. In both cases, it turns out that inertia dominates in the dynamics and keeps the problem well-posed, as we shall see in Secs.~\ref{sec: early phase} and \ref{sec: long sheet}. We therefore retain both inertial and viscous terms in Eq.\,\eqref{eq: dimen mom} 
\footnote{We note that the (arguably non-intuitive)  necessity to retain inertia in the limit $Oh \gg 1$ of highly-viscous liquid is intimately related to the reduced dimensionality of the set-up. Indeed, in an axial geometry (expanding hole), the viscous force, $\partial_x(h\partial_xv)$, in Eq.~\eqref{eq: dimen mom}, is replaced by $\partial_r \left (h r^{-1}  (\partial_r r v_r)\right )$, and a purely viscous solution   is compatible with initial and boundary conditions, regardless of the value of $l_0/h_0$. This subtle relationship between the geometry and the role of inertia appears to have been overlooked in previous studies, e.g. Ref.~\cite{Savva:09}}.
Balancing these contributions yields the visco-inertial length and time scales, respectively given by
\refstepcounter{equation}
$$
l_I = 4\mu\sqrt{\frac{h_0}{\rho\gamma}} \left(= 4 h_0 \text{Oh}\right), \qquad
t_I = \frac{4\mu h_0}{\gamma},
\eqno{(\theequation\mathrm{a},\mathrm{b})}
\label{eq: li and ti}
$$
with $l_I / t_I = u_{\text{TC}}$, the Taylor–Culick speed \eqref{eq: taylor culick speed}. Note that $l_I \gg h_0$ since $\text{Oh}\gg 1$, meaning that inertial effects take place over distances much larger than the initial sheet thickness, consistently with the thin-film approximation.

Lastly, because the film is long but finite, the dynamics also depends on its initial length $l_0$. We therefore introduce the parameter
\begin{equation}
\mathcal{L} = \frac{l_0}{l_I} = \frac{l_0}{4 h_0 \text{Oh}},
\label{eq: l def}
\end{equation}
which, as we shall see, constitutes the single dimensionless group of the problem. Importantly, since both $l_0/h_0$ and $\text{Oh}$ are taken to be large, $\mathcal{L}$ can attain any value.

\subsection{Dimensionless formulation}

We henceforth adopt a dimensionless formulation and introduce the variables
\refstepcounter{equation}
$$
\label{eq: normalization}
X=\frac{x}{l_I},\qquad T = \frac{t}{t_I}, \qquad X_c =\frac{x_c}{l_I}, \qquad H = \frac{h}{h_0},\qquad V = \frac{v}{u_{TC}}.
\eqno{(\theequation\mathrm{a}\!-\!\mathrm{e})}
$$
The dimensionless thickness and velocity profiles are governed by the continuity equation (cf.\,Eq.\,\eqref{eq: continuity})
\begin{equation}
\frac{\partial H}{\partial T} + \frac{\partial}{\partial X}\left(H V\right) = 0,
\label{eq: continuity dim}
\end{equation}
and the momentum equation (cf.\,Eq.\,\eqref{eq: dimen mom})
\begin{equation}
H\frac{D V}{DT} = \frac{\partial}{\partial X}\left(H\frac{\partial V}{\partial X}\right),
\label{eq: momentum dim}
\end{equation}
which are to be solved for $T>0$ in the domain $X_c(T)< X < \mathcal{L}$, subject to the initial conditions (cf.\,Eqs.\,\eqref{eq: init cond dim}-- \eqref{eq: init vel})
\refstepcounter{equation}
$$
\label{eq: initial dim}
H(X,0)=1,\qquad V(X,0) = 0,\qquad X_c(0)=0;
\eqno{(\theequation\mathrm{a},\mathrm{b},\mathrm{c})}
$$
the free-end boundary conditions (cf.\,Eqs.\,\eqref{eq: dimens kine}---\eqref{eq: dimens dyn})
\refstepcounter{equation}
$$
\label{eq: bc free dim}
H\left(X_c(T), T\right)\frac{\partial V}{\partial X}\left(X_c(T), T\right)=-1,\qquad V\left(X_c(T), T\right)= \frac{d X_c}{dT};
\eqno{(\theequation\mathrm{a},\mathrm{b})}
$$
and the fixed-end boundary conditions (cf.\,Eqs. \eqref{eq: no flow cond} --
\eqref{eq: no slope cond}),
\refstepcounter{equation}
$$
\label{eq: bc fixed}
 V\left(\mathcal{L}, T\right) = 0, \qquad \frac{\partial H}{\partial X}\left(\mathcal{L}, T\right) = 0.
\eqno{(\theequation\mathrm{a},\mathrm{b})}
$$

Note that, in accordance with Eq.\,\eqref{eq: capil negl}, capillary terms have been neglected in Eq.\,\eqref{eq: momentum dim}.  Furthermore, we observe that the problem depends on a single parameter $\mathcal{L}$,  defined in Eq.\,\eqref{eq: l def}, which appears in \eqref{eq: bc fixed}.

\section{Problem transformation}
In this section, we show that Eqs.\,\eqref{eq: continuity dim}--\eqref{eq: initial dim} can be recast into a much simpler form. In particular, we demonstrate that the problem possesses a conserved quantity, which enables us to transform Eqs.\,\eqref{eq: continuity dim}--\eqref{eq: momentum dim} into a single equation that depends exclusively on $H$. Alternatively, in Appendix \ref{app: v problem}, we derive a single equation for the problem that depends exclusively on $V$.

\subsection{Conserved quantity}
We begin by showing that the quantity $V + \frac{1}{H}\frac{\partial H}{\partial X}$ is conserved by the flow. Moreover, the initial conditions Eq.\,\eqref{eq: initial dim} imply that
\begin{equation}
V  =  - \frac{1}{H}\frac{\partial H}{\partial X},
\label{eq: h and v rel}
\end{equation}
which relates the fluid velocity to the geometry of the sheet. 

To begin, we rewrite Eq.\,\eqref{eq: continuity dim} as
\begin{equation}
H\frac{\partial V}{\partial X} = -\frac{D H}{DT}.
\label{eq: mass balance 2}
\end{equation}
Substituting the above into Eq.\,\eqref{eq: momentum dim} yields
\begin{equation}
H\frac{D V}{DT} = -\frac{\partial}{\partial X}\left(\frac{D H}{DT}\right),
\label{eq: stress balance extra}
\end{equation}
which, after switching the order of derivatives on the right-hand side, becomes
\begin{equation}
H\frac{D V}{DT} = -\frac{D}{DT}\left(\frac{\partial H}{\partial X}\right) - \frac{\partial V}{\partial X}\frac{\partial H}{\partial X}.
\label{eq: int set con}
\end{equation}
The second term on the right-hand side arises from the convective term $V\frac{\partial}{\partial X}$ appearing in the material derivative. We proceed by dividing both sides of Eq.\,\eqref{eq: int set con} by $H$ and employing the identity
\begin{equation}
\frac{1}{H}\frac{D}{DT}\left(\frac{\partial H}{\partial X}\right) = \frac{D}{DT}\left(\frac{1}{H}\frac{\partial H}{\partial X}\right) + \frac{1}{H^2}\frac{\partial H}{\partial X}\frac{D H}{DT},
\end{equation}
which allows us to rewrite Eq.\,\eqref{eq: int set con} as
\begin{equation}
\frac{D V}{DT} =  - \frac{D}{DT}\left(\frac{1}{H}\frac{\partial H}{\partial X}\right) - \frac{1}{H^2}\frac{\partial H}{\partial X}\left(H\frac{\partial V}{\partial X} + \frac{D H}{DT}\right).
\end{equation}
Since the term $\left(H\frac{\partial V}{\partial X} + \frac{D H}{DT}\right)$ vanishes by the continuity equation Eq.\,\eqref{eq: mass balance 2}, we arrive at the conservation law
\begin{equation}
\frac{D}{DT}\left(V + \frac{1}{H}\frac{\partial H}{\partial X}\right)=0.
\label{eq: conserved}
\end{equation}
Integrating it once using the initial conditions Eq.\,\eqref{eq: initial dim} proves Eq.\,\eqref{eq: h and v rel}.

To physically interpret this result, it is helpful to adopt a Lagrangian description, as done in Appendix~\ref{app: lagrangian}. We note that Eq.\,\eqref{eq: mass balance 2} shows that time variations in film thickness  following a fluid element arise from local stretching. From this, the net viscous force per unit mass acting on the element is $ - \frac{D}{D T} \left( \frac{1}{H} \frac{\partial H}{\partial X} \right)$, as derived in the appendix. Equating this expression to the Lagrangian acceleration $\frac{D V}{D T}$ yields Eq.\,\eqref{eq: conserved}.

\subsection{Reduced problem}
\label{ss: thickness reduced}

Substituting Eq.\,\eqref{eq: h and v rel} into Eq.\,\eqref{eq: continuity dim}, we arrive at the one-dimensional heat equation for $H$,
\begin{equation}
\frac{\partial H}{\partial T} = \frac{\partial^2 H}{\partial X^2}.
\label{eq: diffusion h}
\end{equation}

The thickness profile is subject to the initial conditions (\ref{eq: initial dim}a) and (\ref{eq: initial dim}c), and the zero-slope boundary condition (\ref{eq: bc fixed}b). To close the problem, it remains to rewrite the free-end boundary conditions Eq.\,\eqref{eq: bc free dim} in terms of $H$. To this end, we first substitute Eq.\,\eqref{eq: mass balance 2} into (\ref{eq: bc free dim}a) to obtain
\begin{equation}
\frac{D H}{DT}(X_c(T), T) = 1;
\end{equation}
integrating it once using Eq.\,\eqref{eq: initial dim} furnishes the time-dependent boundary conditions
\refstepcounter{equation}
$$
\label{eq: alternative h cond}
H(X_c(T), T) = 1 + T, \qquad \frac{\partial H}{\partial X}(\mathcal{L}, T) = 0.
\eqno{(\theequation\mathrm{a},\mathrm{b})}
$$
A boundary condition for $X_c(T)$ is obtained by substituting Eqs.\,\eqref{eq: h and v rel} and \eqref{eq: alternative h cond} into Eq.\,(\ref{eq: bc free dim}b):
\begin{equation}
\frac{d X_c}{d T} = -\frac{1}{1 + T}\frac{\partial H}{\partial X}(X_c(T), T).
\label{eq: alternative h cond 2}
\end{equation}

The reduced problem is now closed. Nonetheless, for the analysis in the following sections, it will be convenient to derive an integral mass balance for the problem. To this end, we integrate Eq.\,\eqref{eq: diffusion h} from $X = X_c$ to $X = \mathcal{L}$ using Eqs.\,\eqref{eq: alternative h cond 2} and (\ref{eq: bc fixed}b), which gives
\begin{equation}
\int_{X_c}^{\mathcal{L}} \frac{\partial H}{\partial T} \, dX = \left(1 + T\right)\frac{d X_c}{d T}.
\label{eq: cons1}
\end{equation}
Exchanging the order of the integral and the derivative on the left-hand side, we find
\begin{equation}
\frac{d}{d T} \int_{X_c}^{\mathcal{L}} H(X, T) \, dX + H(X_c, T)\frac{d X_c}{d T} = \left(1 + T\right)\frac{d X_c}{d T}.
\end{equation}
Then, substituting Eq.\,\eqref{eq: alternative h cond} into the second term on the left-hand side shows that $\int_{X_c}^{\mathcal{L}} H \, dX$ is conserved. Accordingly, initial conditions (\ref{eq: initial dim}a) and (\ref{eq: initial dim}c) imply the integral relation
\begin{equation}
\int_{X_c}^{\mathcal{L}} H(X, T) \, dX = \mathcal{L},
\label{eq: mass balance}
\end{equation}
which represents the global mass conservation of the liquid.

We note that our reduced problem is mathematically equivalent to a Stefan problem with time-dependent boundary conditions \cite{Ockendon:03}. This class of problems---first studied in the context of ice formation \cite{Stefan:89}---is characterized by a moving boundary whose position ($X_c(T)$ here) must be determined as part of the solution. While we have been unable to find an exact solution to this problem, we show in Secs.\,\ref{sec: early phase}--\ref{sec: phase changes} that several asymptotic formulae, valid for different stages of retraction, can be derived starting from the reduced problem.

Similar heat-like equations describing the evolution of the thickness profile have been derived previously in the context of gravitationally driven thin-film stretching \cite{kaye:91, stokes:04, Bradshaw-Hajek:07, Wylie:11}. In those studies, the authors also identified a conserved quantity of the flow analogous to Eq.\,\eqref{eq: h and v rel}. However, their analyses considered a different physical setting and employed boundary conditions distinct from those used in the present problem.

\section{Numerical Simulations}

We now present numerical solutions of the reduced problem derived in Sec.\,\ref{ss: thickness reduced}. To facilitate our simulations, we rescale $H$ and $X$ so that the spatial domain and boundary conditions remain time-independent. Details of our numerical scheme are given in Appendix~\ref{app: numerics}.

\begin{figure}[!ht]
\begin{subfigure}{0.45\textwidth}
\caption{$H(X,T),\ \mathcal{L}=0.1$}
\includegraphics[width=\textwidth]{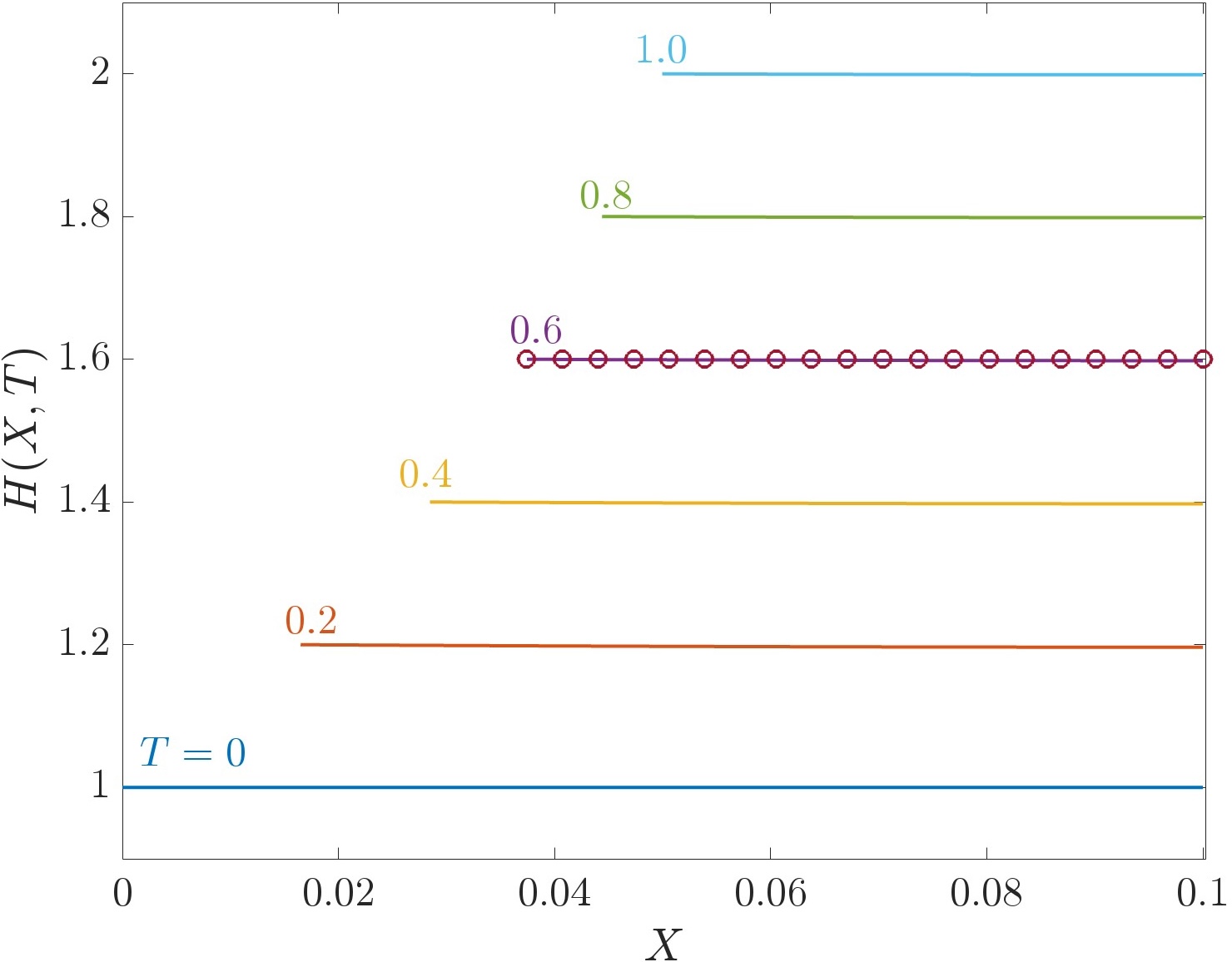}
\label{fig:h_L=0_1}
\end{subfigure}
\begin{subfigure}{0.45\textwidth}
\caption{$V(X,T),\ \mathcal{L}=0.1$}
\includegraphics[width=\textwidth]{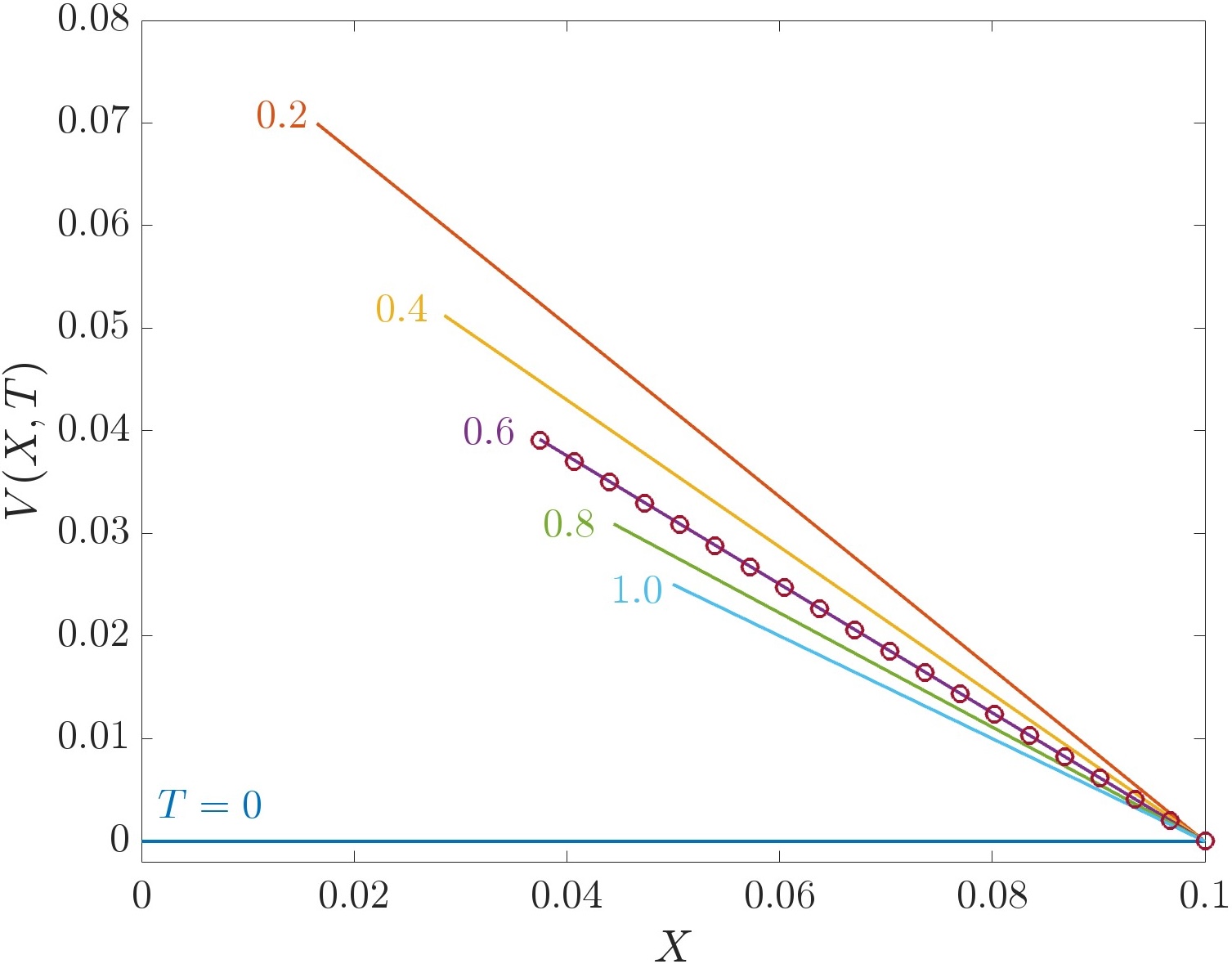}
\label{fig:v_L=0_1}
\end{subfigure}
~
\begin{subfigure}{0.45\textwidth}
\caption{$H(X,T),\ \mathcal{L}=1$}
\includegraphics[width=\textwidth]{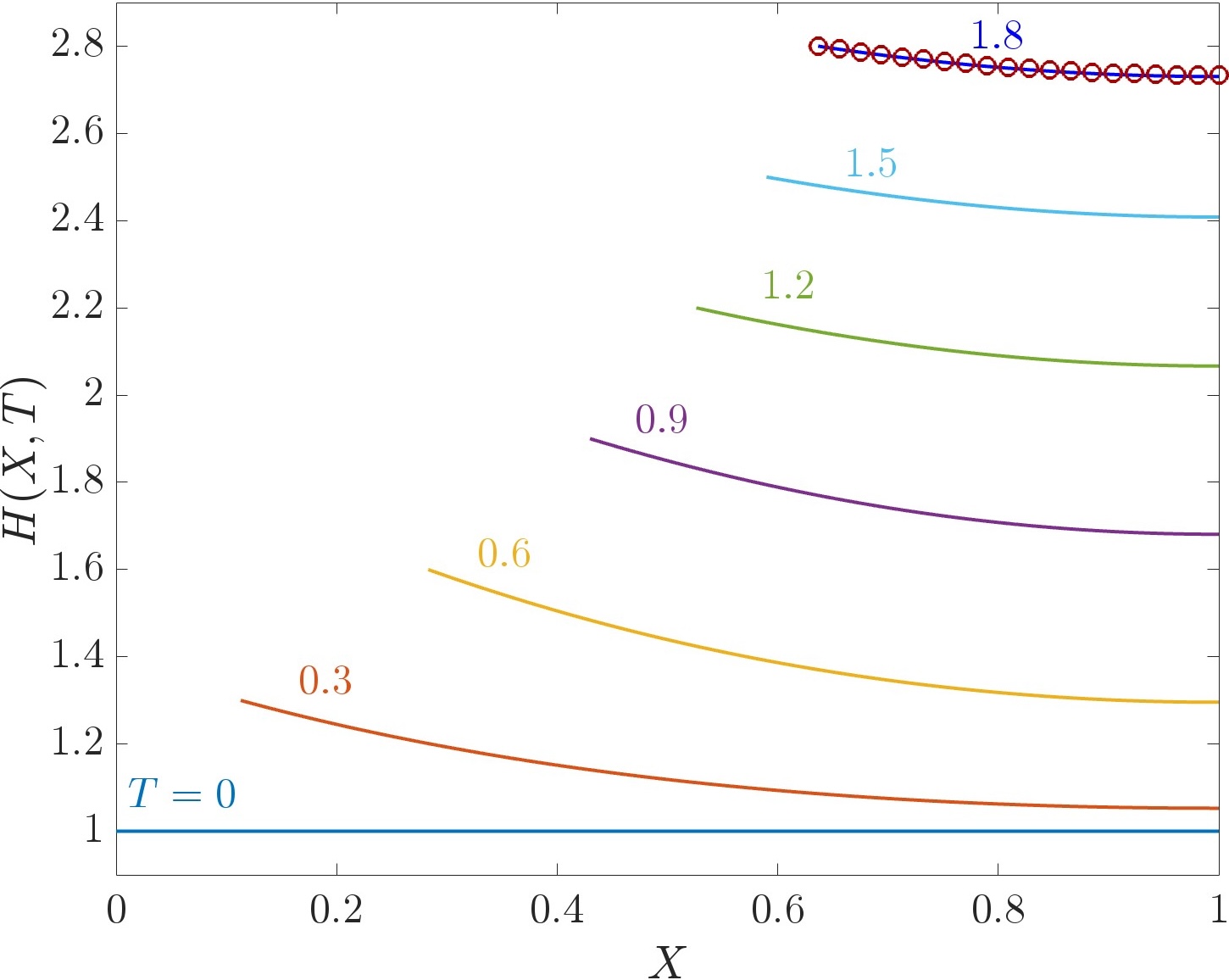}
\label{fig:h_L=1}
\end{subfigure}
\begin{subfigure}{0.45\textwidth}
\caption{$V(X,T),\ \mathcal{L}=1$}
\includegraphics[width=\textwidth]{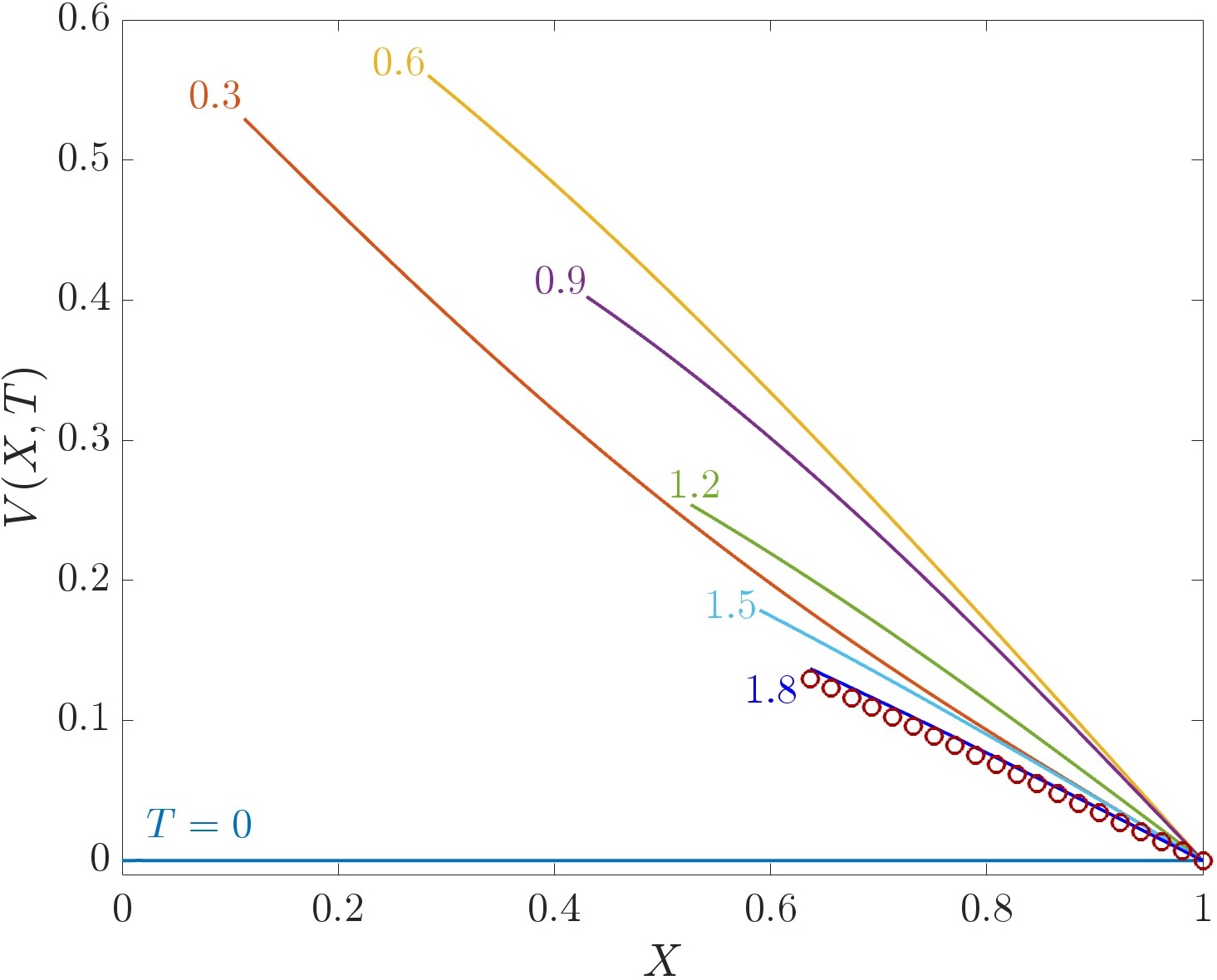}
\label{fig:v_L=1}
\end{subfigure}
~
\begin{subfigure}{0.45\textwidth}
\caption{$H(X,T),\ \mathcal{L}=20$}
\includegraphics[width=\textwidth]{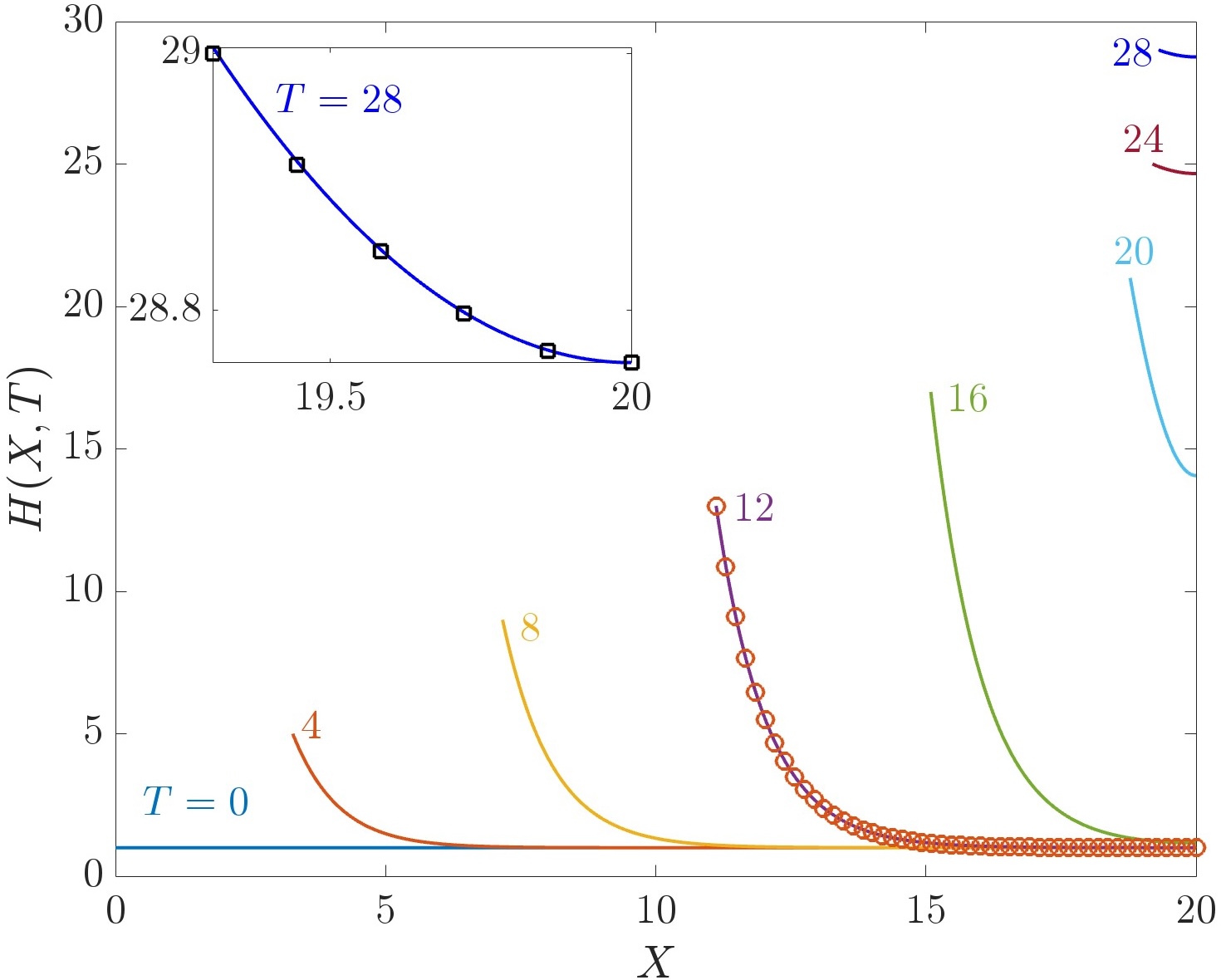}
\label{fig:h_L=20}
\end{subfigure}
\begin{subfigure}{0.45\textwidth}
\caption{$V(X,T),\ \mathcal{L}=20$}
\includegraphics[width=\textwidth]{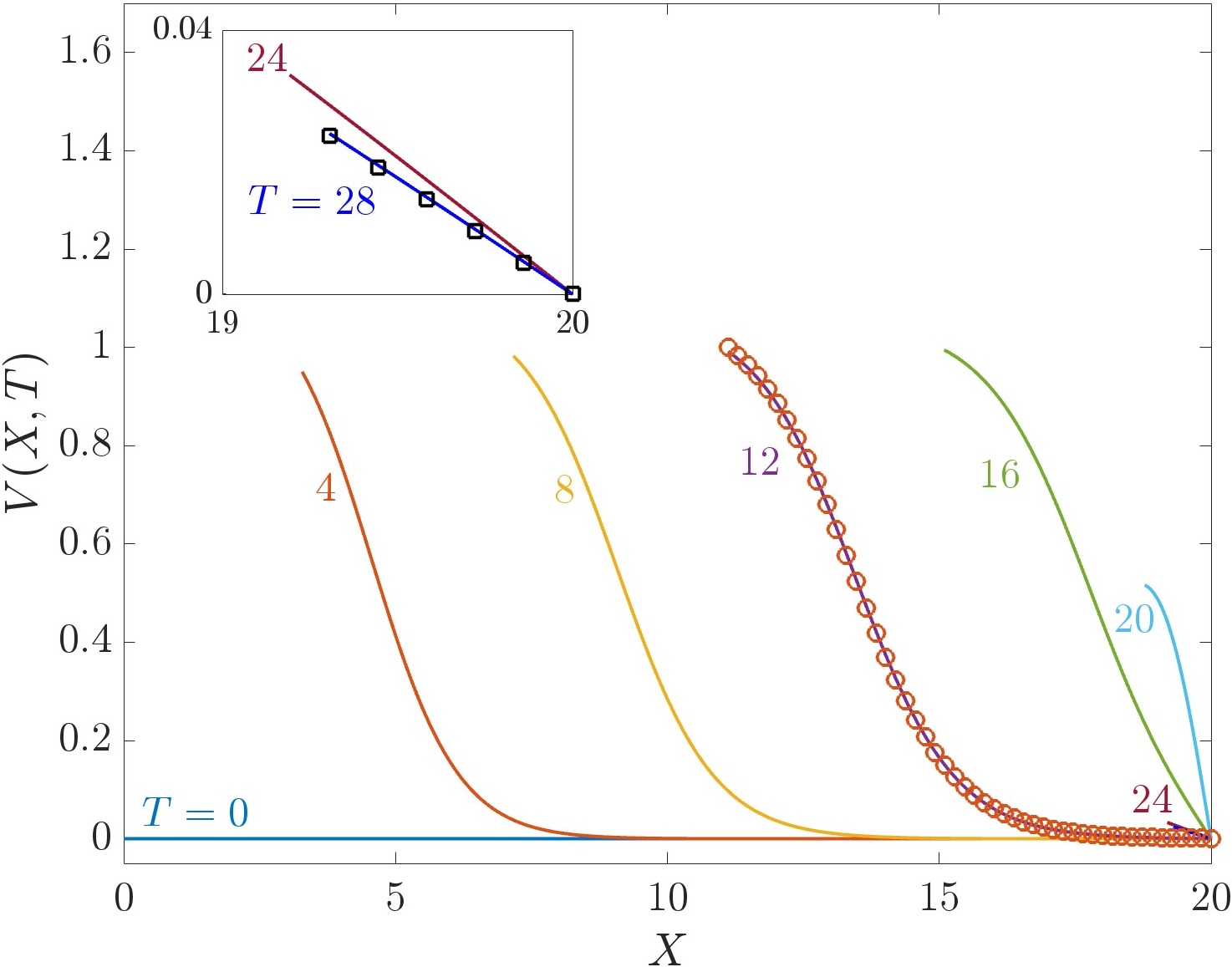}
\label{fig:v_L=20}
\end{subfigure}
\caption{Numerically computed profiles of the thickness $H(X,T)$ and velocity $V(X,T)$ are shown for $\mathcal{L} = 0.1$, $\mathcal{L} = 1$, and $\mathcal{L} = 20$ at various times. In panels (a–d), the circles indicate the asymptotic predictions for $H$ and $V$ given by Eqs.\,\eqref{eq: H short expansion} and \eqref{eq: V asympt short}. In panels (e) and (f), the circles correspond to the asymptotic formulae in Eqs.\,\eqref{eq: wkb text} and \eqref{eq: velocity profile}, while the squares in the insets show the results from Eqs.\,\eqref{eq: H short expansion} and \eqref{eq: V asympt short}.}
\label{fig:profile}
\end{figure}

We begin by showing  the thickness and velocity profiles, $H(X,T)$ and $V(X,T)$, for various times in Fig.~\ref{fig:profile}. The velocity $V$ is obtained from $H$ via Eq.\,\eqref{eq: h and v rel}. To illustrate the impact of $\mathcal{L}$ on the dynamics, each row of Fig.~\ref{fig:profile} corresponds to a different value of the dimensionless length $\mathcal{L}$, specifically $\mathcal{L} = 0.1$, $1$, and $20$.

For the shortest sheet with $\mathcal{L} = 0.1$ (Figs.~\ref{fig:profile}(a) and (b)), the sheet thickness remains nearly uniform across the domain, gradually increasing over time.  The velocity profile varies approximately linearly with $X$, vanishing at the fixed end 
$X = \mathcal{L}$
and attaining its maximum at the free edge $X = X_c$, in accordance with the boundary condition Eq.\,(\ref{eq: bc fixed}a). However, the film is not exactly flat; otherwise, Eq.\,\eqref{eq: h and v rel} would imply $V=0$ everywhere.

For $\mathcal{L} = 1$ (Figs.~\ref{fig:profile}(c) and (d)), the thickness profile exhibits more noticeable spatial variation, with the film thickest at $X = X_c$ and gradually thinning toward the opposite end. These variations are most pronounced at intermediate times ($T = 0.6$ and $T \approx 0.8$) and diminish over time as the film becomes more spatially uniform, approaching the behavior observed for $\mathcal{L} = 0.1$. The velocity profile shows a nonlinear dependence on $X$ at intermediate times, which gradually transitions to a linear variation at later stages.

We observe similar trends for $\mathcal{L} = 20$ (Figs.~\ref{fig:profile}(e) and (f)), where the nonlinear dependence of both $H$ and $V$ on $X$ becomes even more pronounced. A distinctive feature, not seen for smaller $\mathcal{L}$, is that the velocity profile remains nearly unchanged from $T = 4$ to $T = 20$, starting at approximately unity at $X = X_c$ and decreasing to zero at $X = \mathcal{L}$. As in the previous cases, the film thickness gradually becomes more uniform over time. This is evident in the inset of Fig.~\ref{fig:profile}(e), which shows that by $T = 28$, the relative difference between the maximum and minimum thickness ($H \approx 29$ and $28.8$, respectively) is less than 1\%. Similarly, the inset of Fig.~\ref{fig:profile}(f) shows that for large $T$, the velocity profile becomes nearly linear in $X$.

\begin{figure}[!ht]
\begin{subfigure}{0.46\textwidth}
\includegraphics[width=\textwidth]{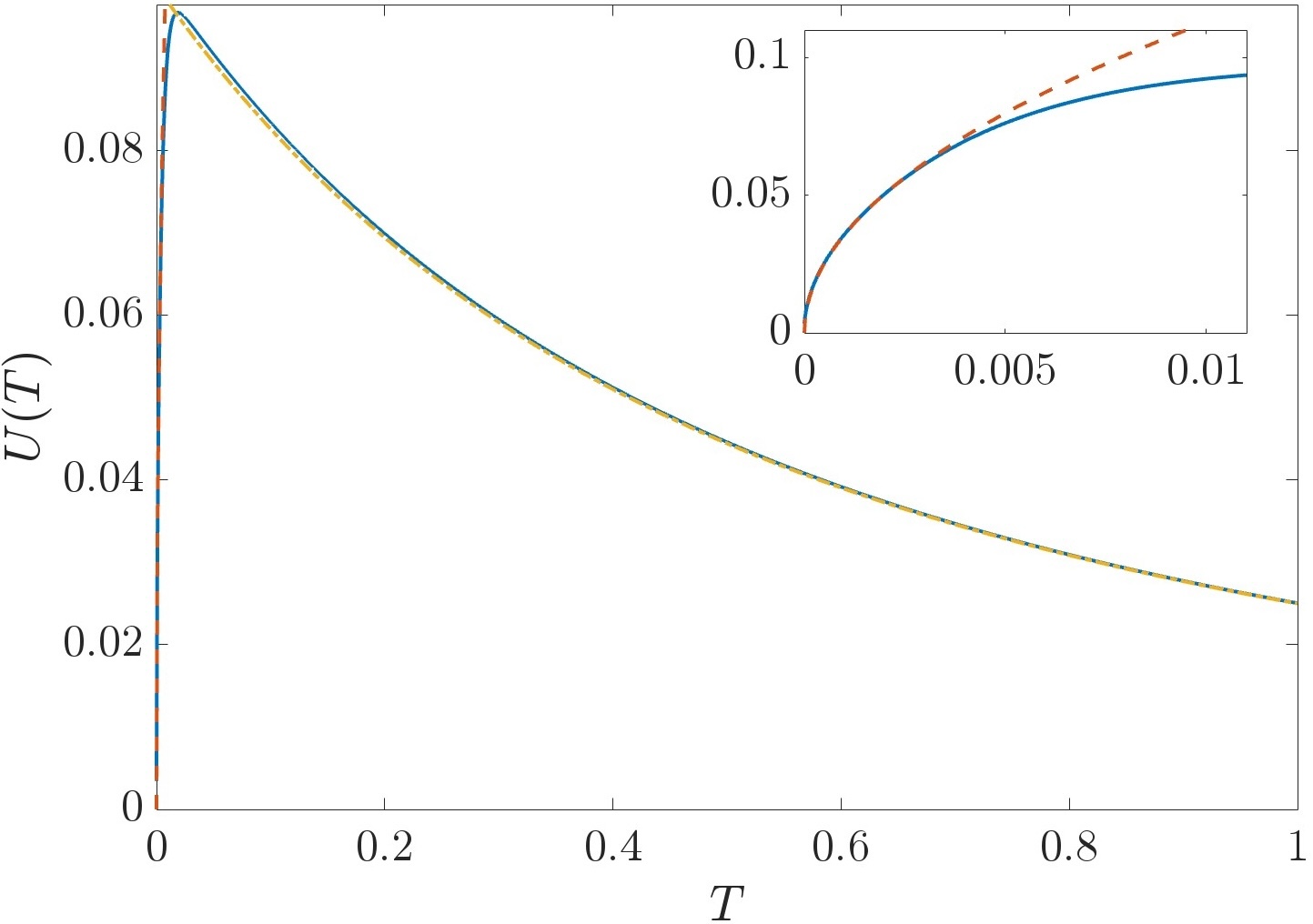}
\caption{$\mathcal{L}=0.1$}
\end{subfigure}
~
\begin{subfigure}{0.45\textwidth}
\includegraphics[width=\textwidth]{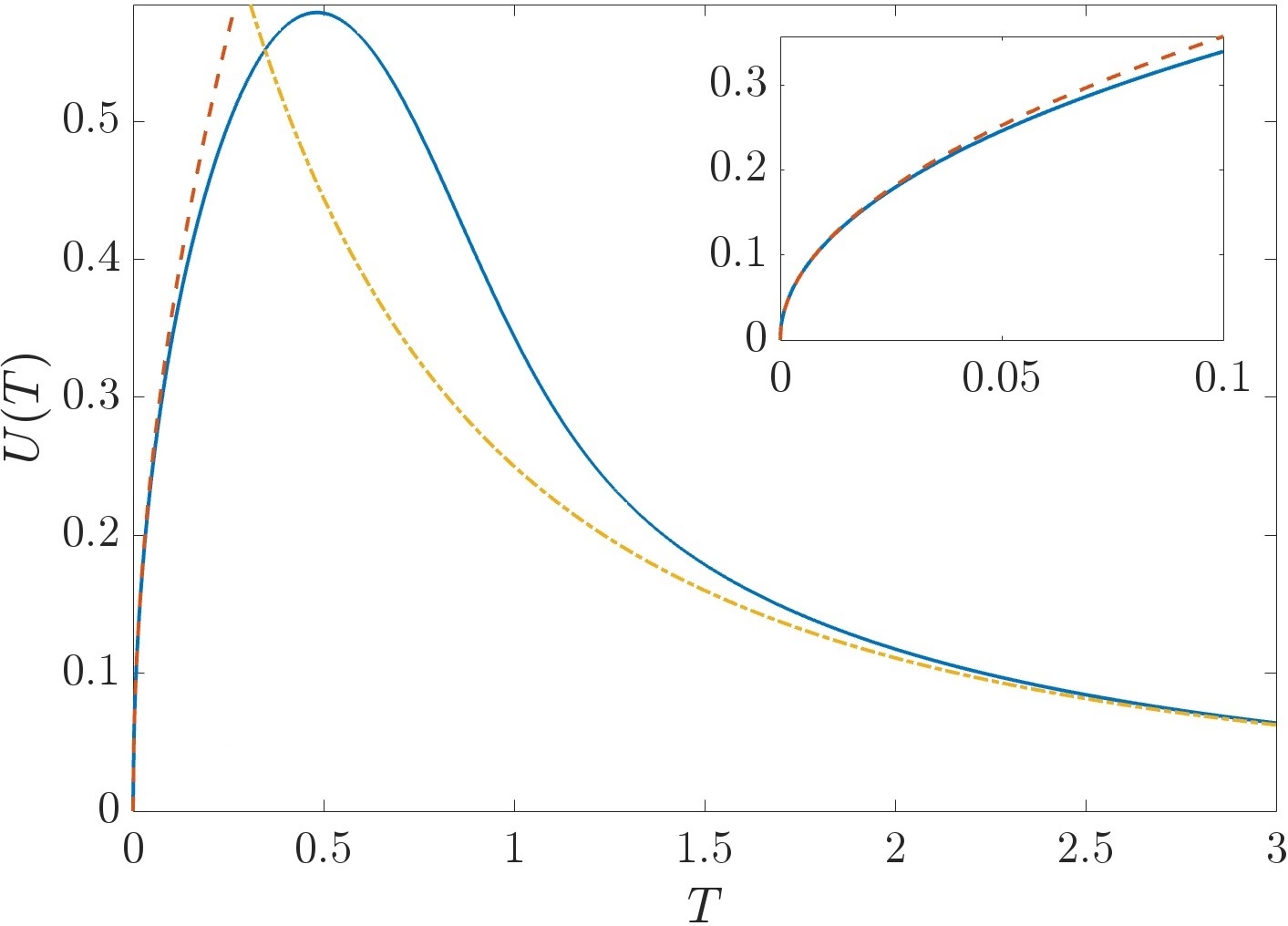}
\caption{$\mathcal{L}=1$}
\end{subfigure}

\begin{subfigure}{0.6\textwidth}
\centering
\includegraphics[width=\textwidth]{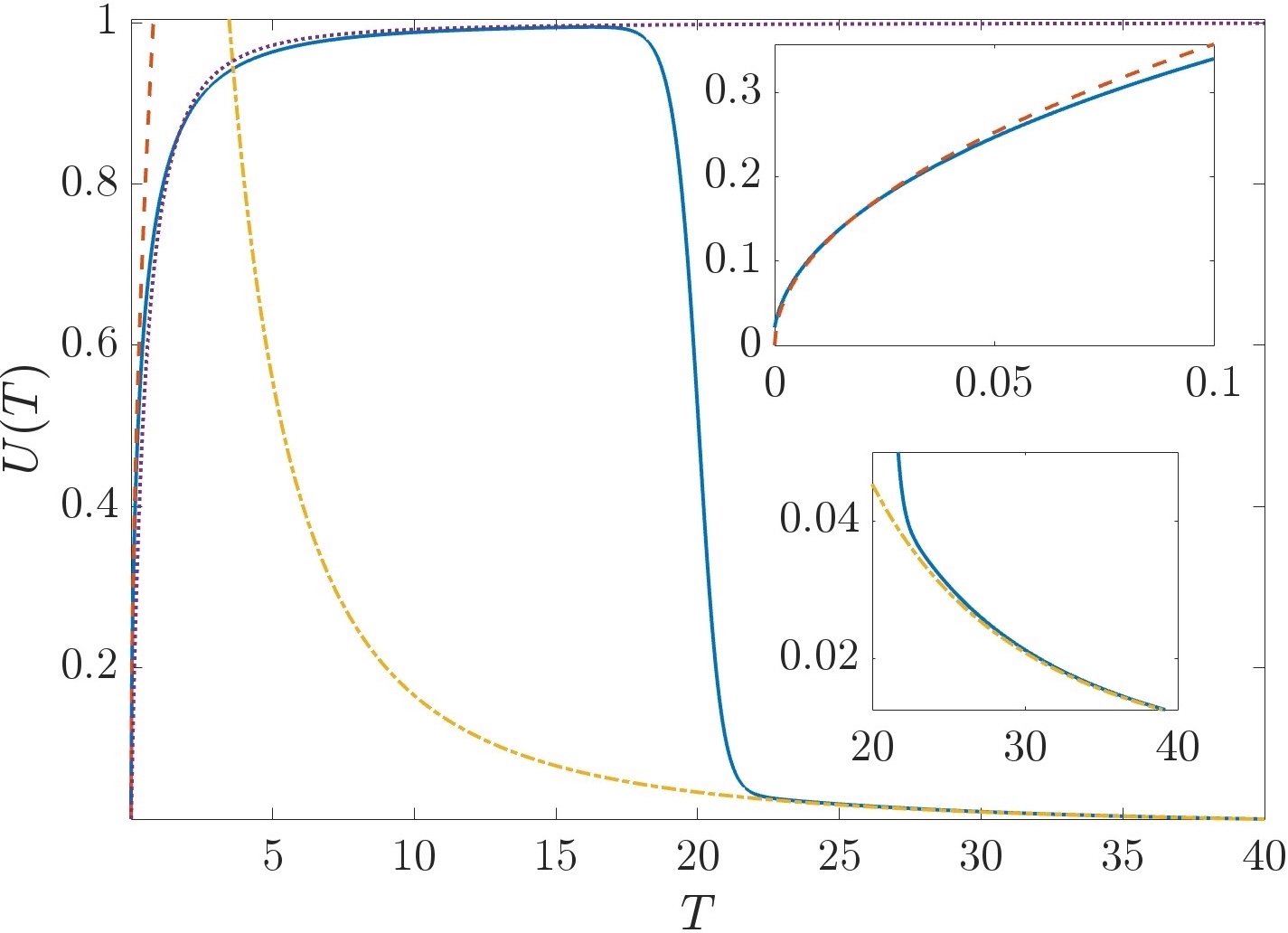}
\caption{$\mathcal{L}=20$}
\end{subfigure}
\caption{Time evolution (blue line) of the retraction speed $d X_c/dT$ for (a) $\mathcal{L}=0.1$, (b) $\mathcal{L}= 1$,  and (c) $\mathcal{L}=20$.  The dashed red line is the early-time asymptotic Eq.\,\eqref{eq: xc asympt small t}. The yellow dotted-dashed line is the late-time asymptotics Eq.\,\eqref{eq: u asymp short}. The purple dotted line in (c) is the infinite-sheet asymptotics Eq.\,\eqref{eq: infinite asymptotics}.}
\label{fig: u_time}
\end{figure}

To further explore the retraction dynamics, Fig.\,\ref{fig: u_time} shows the time evolution of the retraction speed, $U(T) = dX_c/dT$, for $\mathcal{L} = 0.1$, $1$, and $20$. In all cases, the speed starts at zero, rises to a peak at a time $T = T_{\text{max}}$, and then decreases monotonically, eventually tending to zero as $T \to \infty$. Log-log plots reveal consistent power-law behavior for all $\mathcal{L}$: at early times, the retraction speed scales as $T^{1/2}$, while at late times it decays as $1/T^2$.

\begin{figure}[!t]
\centering
\includegraphics[width=0.6\textwidth]{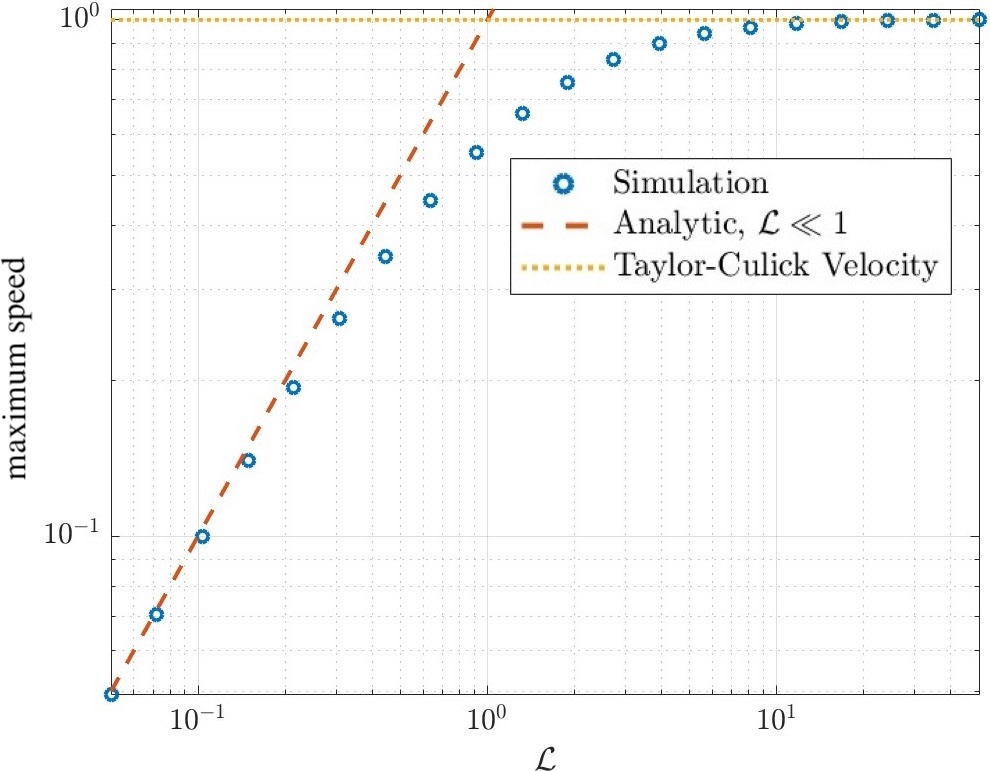}
\caption{The maximum value of $d X_c/dT$ reached by the thin film during its retraction for different $\mathcal{L}$. The dashed red line is the maximum of Eq.\,\eqref{eq: u asymp short}, which is simply given by $\mathcal{L}$, and the dotted yellow line is the Taylor-Culick velocity \eqref{eq: taylor culick speed}.}
\label{fig: u_max}
\end{figure}

Notably, the maximum retraction speed is always below the Taylor–Culick speed, given by Eq.\,\eqref{eq: taylor culick speed}, which equals unity in dimensionless form. For $\mathcal{L} = 0.1$ and $\mathcal{L} = 1$ in Figs.\,\ref{fig: u_time}(a,b), the maximum speed is substantially below unity. In contrast, for the longest sheet ($\mathcal{L} = 20$, Fig.\,\ref{fig: u_time}(c)), the retraction speed approaches the Taylor–Culick value, remaining close to it for a finite interval before undergoing a sharp deceleration near $T \approx \mathcal{L}$. In addition, 
Fig.\,\ref{fig: u_max} further shows how the peak retraction speed varies with $\mathcal{L}$ (blue circles) and confirms that the Taylor–Culick speed acts as an upper bound, with the actual maximum speed asymptotically approaching this limit as $\mathcal{L}$ increases.

In summary, our numerics show that for all $\mathcal{L}$ the retraction dynamics follow similar trends at early and late times. At early times, the retraction speed grows as $T^{1/2}$; at late times, it decreases as $1/T^2$, accompanied by a nearly uniform thickness profile and a linearly varying velocity profile. However, at intermediate times, the dynamics depend on $\mathcal{L}$, with the maximum speed increasing as $\mathcal{L}$ grows. For large $\mathcal{L}$, a distinct intermediate regime emerges, where the speed remains close to the Taylor-Culick value over a finite time interval.

We also note that numerical solutions are in good agreement with the full Navier–Stokes simulations in \cite{Deka:20}, which examined the influence of $\text{Oh}$ and $l_0/h_0$ on the retraction of a viscous sheet. In particular, the retraction speed $U(T)$ reported for $\text{Oh} = 20$ and $l_0/h_0 = 80$ ($\mathcal{L} = 1$), shown in Fig. 7 of that paper, closely matches our result in Fig.\,\ref{fig: u_time}(b), with both exhibiting a peak speed of approximately $U(T)\approx 0.55$ at $T \approx 0.6$.

To gain further insight into our numerical findings,  in the following sections we derive asymptotic formulae for the retraction speed, as well as the thickness and velocity profiles, during different stages of retraction and for both small and large values of $\mathcal{L}$.

\section{Early Times $(T\ll 1)$}
\label{sec: early phase}

We first derive asymptotic approximations valid during the early stages of retraction, corresponding to the limit $T\ll1$ with $\mathcal{L}$ fixed. The thickness profile $H$ is governed by the heat equation Eq.\,\eqref{eq: diffusion h}, which suggests that variations in $H$ initially occur over the characteristic diffusive length scale of order $\sqrt{T}$. Furthermore,  Eq.\,\eqref{eq: alternative h cond} suggests that the thickness variation $H-1 = \mathcal{O}(T)$. With this in mind, we introduce the self-similar variable $\omega = X/\sqrt{T}$ and pose the two-term expansion
\begin{equation}
H \sim 1 + T\tilde{H}\left(\omega\right) \quad\text{as}\quad T\to0.
\label{eq: H small T expansion}
\end{equation}

In terms of $\omega$, the fixed end is located at $\omega = \mathcal{L}/T^{1/2}$, such that $\omega\to\infty$ as $T\to0$. The free edge is located at $\omega_c = X_c(T)/T^{1/2}$. To calculate $\omega_c$ as $T\to0$, we must first estimate $X_c(T)$. Expanding Eq.\,\eqref{eq: alternative h cond}, we find
\begin{equation}
\label{eq: early T exp}
\frac{d X_c}{dT} \sim - T^{1/2}\frac{d\tilde{H}}{d\omega},
\end{equation}
which can be integrated in time to show that $X_c(T) = \mathcal{O}(T^{3/2})$. Hence, $\omega_c = X_c(T)/\sqrt{T}\to0$ as $T\to0$. Hence, in the $\omega$-coordinate, the horizontal domain becomes the semi-infinite interval 
\begin{equation}
0<\omega<\infty.
\end{equation}

We proceed by expanding Eq.\,\eqref{eq: diffusion h} using Eq.\,\eqref{eq: H small T expansion}. To leading order, we arrive at a second-order ordinary differential equation
\begin{equation}
\tilde{H} - \frac{1}{2}\omega \frac{ d\tilde{H}}{d \omega} = \frac{d^2 \tilde{H}}{d\omega^2}.
\label{eq: diff self sim}
\end{equation}
Similarly, expanding the boundary conditions at the free edge, Eq.\,\eqref{eq: alternative h cond}, and fixed end, Eq.\,(\ref{eq: initial dim}a), we find, to leading order,
\begin{equation}
\label{eq: initial condition-zero}
\tilde{H}(\omega=0)=1
\end{equation}
and
\begin{equation}
\label{eq: initial condition}
\tilde{H}(\omega) = o(\omega^2)\quad \text{as}\quad\omega\to\infty. 
\end{equation}
The solution of Eq.\,\eqref{eq: diff self sim} subject to Eqs.\,\eqref{eq: initial condition-zero} and \eqref{eq: initial condition} is
\begin{equation}
\tilde{H}(\omega) =   \frac{1}{2}\left(2 + \omega^2\right) \text{erfc}\left(\frac{\omega}{2}\right)-\frac{\omega}{\sqrt{\pi}}\exp\left(-\frac{\omega^2}{4}\right),
\end{equation}
where $\text{erfc}(.)$ is the complementary error function \cite{Abramowitz:book}. In terms of the original variables, the early-time behavior Eq.\,\eqref{eq: H small T expansion} becomes
\begin{equation}
H \sim 1 + \frac{1}{2}\left(2 T + X^2\right) \text{erfc}\left(\frac{X}{2 \sqrt{T}}\right)-\frac{X}{\sqrt{\pi T}}\exp\left(-\frac{X^2}{4 T}\right).
\label{eq: H approx early}
\end{equation}
From Eq.\,\eqref{eq: h and v rel}, the early-time behavior of the velocity profile is
\begin{equation}
V \sim  2\sqrt{\frac{T}{\pi}}\exp\left(-\frac{X^2}{4 T}\right)  -\frac{X}{\sqrt{T}}\text{erfc}\left(\frac{X}{2 \sqrt{T}}\right),
\label{eq: v asympt small t}
\end{equation}
and, from Eq.\,\eqref{eq: alternative h cond}, the early-time behavior of the retraction speed is 
\begin{equation}
\frac{d X_c}{dT} \sim 2\sqrt{\frac{T}{\pi}}.
\label{eq: xc asympt small t}
\end{equation}
Expressions identical to Eqs.\,\eqref{eq: v asympt small t} and Eq.\,\eqref{eq: xc asympt small t} were previously derived in \cite{Savva:09} using different arguments. The early-time approximation for the retraction speed Eq.\,\eqref{eq: xc asympt small t} is depicted by the dashed red curves in Fig.\,\ref{fig: u_time}.

\section{Late times: finite films $\left (\mathcal{L}- X_c\ll 1\right )$}
\label{sec: late phase}

The simulations show that the sheet shortens as it retracts, with $X_c \to \mathcal{L}$ as $T\to\infty$. This motivates an asymptotic analysis in the limit $\mathcal{L}-X_c\ll 1$, which describes the late-time dynamics of the sheet. Importantly, the limit $\mathcal{L}-X_c\ll 1$ is more general than $T\gg 1$. For instance, if $\mathcal{L}\ll1$, then $\mathcal{L}-X_c\ll 1$ holds even for $T\ll1$. Accordingly, here we analyze the regime $\mathcal{L}-X_c\ll 1$ while keeping $T$ fixed. In dimensional terms, this corresponds to $l_0 - x_c \ll l_I$, where $l_I$ is the visco-inertial length scale defined in Eq.\,(\ref{eq: li and ti}a), representing a regime in which the sheet is sufficiently short for viscous forces to dominate over inertia.

Since $\mathcal{L}-X_c$ is small, the spatial derivative dominates in Eq.\,\eqref{eq: diffusion h}, and, at leading order, we obtain
\begin{equation}
\frac{\partial^2 H}{\partial X^2} \approx 0.
\end{equation}
Integrating twice using Eq.\,\eqref{eq: alternative h cond} and Eq.\,\eqref{eq: alternative h cond 2}, we find
\begin{equation}
H \sim 1 + T.
\label{eq: leading H short expansion}
\end{equation}

Since the leading-order thickness is uniform, we cannot determine $V$ from Eq.\,\eqref{eq: h and v rel}. To resolve this, we proceed to the next order and pose the expansion
\begin{equation}
H \sim 1 + T + \delta(X,T),
\label{eq: H short expansion}
\end{equation}
as $X_c \to \mathcal{L}$, where the correction $\delta \to 0$ in the indicated limit. Substituting into Eqs.\,\eqref{eq: diffusion h}, \eqref{eq: alternative h cond}, and \eqref{eq: bc fixed}, we find that $\delta$ satisfies
\begin{gather}
\frac{\partial^2 \delta }{\partial X^2} = 1,
\end{gather}
subject to the boundary conditions
\refstepcounter{equation}
$$
\delta  = 0 \quad \text{at}\quad X=X_c, \qquad \frac{\partial{\delta}}{\partial X}=0\quad \text{at}\quad X=\mathcal{L}.
\eqno{(\theequation\mathrm{a},\mathrm{b})}
$$
The solution is
\begin{equation}
\delta = \frac{1}{2}\left(X + X_c - 2\mathcal{L}\right)\left(X - X_c\right).
\label{eq: H asympt short}
\end{equation}
Since $X_c\le X\le \mathcal L$, we have $|\delta|=\tfrac12(2\mathcal L-X-X_c)(X-X_c)\le (\mathcal L-X_c)^2$. Thus $\delta\ll 1$ for $\mathcal L-X_c\ll 1$, and $\delta$ is asymptotically smaller than the leading-order term \eqref{eq: leading H short expansion}.
Substituting into Eq.\,\eqref{eq: h and v rel}, we obtain the velocity profile
\begin{equation}
V(X,T) \sim \frac{\mathcal{L} - X}{1 + T}.
\label{eq: V asympt short}
\end{equation}

To determine the retraction speed, we substitute Eqs.\,\eqref{eq: H short expansion} and \eqref{eq: H asympt short} into Eq.\,\eqref{eq: alternative h cond}, which at leading order gives
\begin{equation}
\frac{dX_c}{dT} \sim \frac{\mathcal{L} - X_c}{1 + T}.
\label{eq: Xcdt asympt short}
\end{equation}
Solving Eq.\,\eqref{eq: Xcdt asympt short} yields
\begin{equation}
X_c(T) = \frac{\mathcal{L} T + c}{1 + T},
\label{eq: X_c asympt short}
\end{equation}
where $c$ is a constant of integration. Using mass conservation Eq.\,\eqref{eq: mass balance}, we find $c = 0$. Substituting Eq.\,\eqref{eq: X_c asympt short} into Eq.\,\eqref{eq: H asympt short} furnishes the retraction speed in terms of $\mathcal{L}$ and $T$,
\begin{equation}
\frac{dX_c}{dT} \sim \frac{\mathcal{L}}{\left(1 + T\right)^2}.
\label{eq: u asymp short}
\end{equation}
The late-time approximation for the retraction speed is depicted by the dash-dotted yellow curves in Fig.\,\ref{fig: u_time}.

\section{Late times: infinitely-long sheets  $\left (\mathcal{L}=\infty, \, T\gg 1\right )$}
\label{sec: long sheet}

For sufficiently long sheets ($\mathcal{L} \gg 1$) our simulations reveal an intermediate stage during which the retraction speed saturates near the Taylor–Culick value, $\frac{dX_c}{dT} \sim 1$. This regime arises only if the sheet remains sufficiently long, $\mathcal{L} - X_c \gg 1$, after a sufficiently long time has passed since rupture. To better understand it, here we examine the dynamics of an infinitely long sheet $\mathcal{L}=\infty$ in the limit $T \gg 1$. The regime of validity of this limit for large but finite $\mathcal{L}$ is discussed in Sec.\,\ref{sec: phase changes}.

We begin by taking $\mathcal{L} = \infty$. Since the film is now infinitely long, it is convenient to introduce a shifted coordinate system centered at the free edge,
\begin{equation}
\hat{X} = X - X_c.
\end{equation}
In this new frame, the fluid occupies the semi-infinite domain $0 < \hat{X} < \infty$. 

\subsection{Thickness profile}

Rewriting Eq.\,\eqref{eq: diffusion h} in terms of $\hat{X}$, we obtain
\begin{equation}
\frac{\partial H}{\partial T} - \frac{dX_c}{dT} \frac{\partial H}{\partial \hat{X}} = \frac{\partial^2 H}{\partial \hat{X}^2},
\label{eq: diffusion h long}
\end{equation}
where the advection term accounts for the velocity of the moving frame.

Far from the moving edge, as $\hat{X} \to \infty$, the boundary condition Eq.\,\eqref{eq: alternative h cond 2} implies
\begin{equation}
\frac{\partial H}{\partial \hat{X}} \to 0 \quad \text{as} \quad \hat{X} \to \infty.
\label{eq: long sheet bc}
\end{equation}
Integrating once, we find that $H$ tends to a constant at large distances. By causality, this limiting value must match the initial sheet thickness. Thus, Eq.\,\eqref{eq: long sheet bc} is equivalent to
\begin{equation}
H \to 1 \quad \text{as} \quad \hat{X} \to \infty.
\label{eq: long sheet bc 2}
\end{equation}

We now take the limit $T \gg 1$ and expand both $H$ and $\frac{dX_c}{dT}$ in powers of  $1/T$. The boundary conditions in Eq.\,\eqref{eq: alternative h cond} suggest that $H$ scales as $T$, while $\frac{dX_c}{dT}$ remains of order unity. This motivates the expansions 
\begin{subequations}
\begin{align}
H(X,T) &= T H_1(\hat{X}) + H_0(\hat{X}) + \frac{1}{T}H_{-1}(\hat{X}) + \mathcal{O}\left(\frac{1}{T}\right), \\
\frac{dX_c}{dT} &= U_0 + \frac{U_{-1}}{T} + \frac{U_{-2}}{T^2} + \mathcal{O}\left(\frac{1}{T^2}\right).
\end{align}
\label{eq: h_long_exp}
\end{subequations}

By expanding Eq.\,\eqref{eq: diffusion h long} up to $1/T)$, we obtain the system of equations:
\begin{subequations} \begin{align} \frac{d^2 H_1}{d\hat{X}^2} + U_0 \frac{dH_1}{d\hat{X}} &= 0,\\\frac{d^2 H_0}{d\hat{X}^2} + U_0 \frac{dH_0}{d\hat{X}} &= H_1 - U_{-1} \frac{dH_1}{d\hat{X}}, \\ \frac{d^2 H_{-1}}{d\hat{X}^2} + U_0 \frac{dH_{-1}}{d\hat{X}} &= -U_{-1} \frac{dH_0}{d\hat{X}} - U_{-2} \frac{dH_1}{d\hat{X}}. \end{align} \label{eq: H_subequations} \end{subequations}

Doing the same for the boundary conditions in Eqs.\,\eqref{eq: alternative h cond} and \eqref{eq: long sheet bc 2}, we obtain the boundary and far-field conditions  
\begin{subequations}
\begin{align}
H_1 &= 1 \quad \text{at} \quad \hat{X} = 0, \qquad H_1 \to 0 \quad \text{as} \quad \hat{X} \to \infty, \\
H_0 &= 1 \quad \text{at} \quad \hat{X} = 0, \qquad H_0 \to 1 \quad \text{as} \quad \hat{X} \to \infty, \\
H_{-1} &= 0 \quad \text{at} \quad \hat{X} = 0, \qquad H_{-1} \to 0 \quad \text{as} \quad \hat{X} \to \infty.
\end{align}
\label{eq: BC_asymptotic_2}
\end{subequations}
Solving the equations for $H_1$, $H_0$, and $H_{-1}$ order-by-order yields
\begin{subequations}
\begin{align}
H_1 &= \exp\left(-U_0 \hat{X}\right), \\
H_0 &= 1 - \frac{1 + U_0 U_{-1}}{U_0} \hat{X} \exp\left(-U_0 \hat{X}\right), \\
H_{-1} &= -U_{-2} \hat{X} \exp\left(-U_0 \hat{X}\right) + \frac{U_{-1}}{U_0} \left(U_{-1} U_0 - 1\right) \hat{X}^2 \exp\left(-U_0 \hat{X}\right).
\end{align}
\label{eq: sol_asymptotic}
\end{subequations}

To determine the coefficient $U_0$, $U_{-1}$, $U_{-2}$, we now apply Eq.\,\eqref{eq: alternative h cond 2}. Expanding that equation up to $\mathcal{O}(1/T)$, we arrive at the relations
\refstepcounter{equation}
$$
\label{eq: bc_asymptotic 3}
U_0 = -\frac{d H_{1}}{d\hat{X}},\qquad U_{-1} = -\frac{d H_{0}}{d\hat{X}} + \frac{d H_{1}}{d\hat{X}} ,\qquad U_{-2} = -\frac{d H_{-1}}{d\hat{X}} + \frac{d H_{0}}{d\hat{X}} -\frac{d H_{1}}{d\hat{X}} \quad \text{at}\quad\hat{X}=0.
\eqno{(\theequation\mathrm{a}\!-\!\mathrm{c})}
$$
Substituting \eqref{eq: sol_asymptotic} into these relations, we find
\begin{equation}
U_0 = 1, \qquad U_{-1} = 0.
\label{eq: vel expressions}
\end{equation}
Rewritten in dimensional form, our results show that the retraction speed approaches the Taylor–Culick speed \eqref{eq: taylor culick speed} for $T \gg 1$, provided the sheet is initially sufficiently long. Interestingly, this derivation is fundamentally different from the classical analyses by Taylor and Culick \cite{Taylor:59, Culick:60}, which addressed the opposite limit, $Oh \ll 1$; it shows that when $\text{Oh}$, the free edge reaches this speed even before a circular rim can form.

At this stage, the coefficient $U_{-2}$ remains undetermined. We have confirmed that corrections of $O(1/T^2)$ and beyond cannot be obtained by extending our analysis to higher orders in $1/T$. We believe these corrections depend on the full history of the retraction, and hence we do not attempt to compute them explicitly. 
Instead, we compare our asymptotic predictions with the numerics in Fig.\,\ref{fig: u_time}(c). This comparison confirms that $\frac{dX_c}{dT} - 1 = O(1/T^2)$. Moreover, a particularly good fit to the numerical data is provided by
\begin{equation}\label{eq: infinite asymptotics}
\frac{dX_c}{dT} \sim 1 - \frac{1}{(1+T)^2}\ ,
\end{equation}
which suggests that for the initial conditions in our numerical solution, $U_{-2} = -1$.

We now briefly comment on the uniformity of the asymptotic expansion in Eqs.\,\eqref{eq: h_long_exp} and \eqref{eq: vel expressions}. Retaining all terms up to order unity in $H$, we have
\begin{equation}
H= \underbrace{T\exp\left(-\hat{X}\right)}_{O(T)} + \underbrace{1 - \hat{X}\exp\left(-\hat{X}\right)}_{O(1)} + \mathcal{O}\left(\frac{1}{T}\right).
\label{eq: expansion nonuni}
\end{equation}
We observe that the asymptotic ordering changes as $\hat{X}$ increases. For $\hat{X} = \ln T$, the leading-order term becomes $T\exp\left(-\hat{X}\right) = 1$, and is thus comparable to the second term in Eq.\,\eqref{eq: expansion nonuni}. Furthermore, for $\hat{X} = O(T)$, the two exponential terms, $T\exp\left(-\hat{X}\right)$ and $\hat{X}\exp\left(-\hat{X}\right)$, are of the same order. This non-uniformity in Eq.\,\eqref{eq: expansion nonuni} is undesirable and can be resolved using multiple-scale methods. In Appendix \ref{app: velocity alternative}, we employ the Wentzel–Kramers–Brillouin (WKB) method \cite{Hinch:book} to derive the two-term approximation
\begin{equation}
H \sim 1 + T\exp\left(-\frac{1+T}{T}\hat{X}\right),
\label{eq: wkb text}
\end{equation}
which recovers the first two terms in the expansion for $H$,  and is uniformly valid for $\hat{X} = O(T)$.  

\subsection{Velocity profile}

As before, approximations for the velocity profile can be obtained using \eqref{eq: h and v rel}. In particular, substituting \eqref{eq: wkb text} yields
\begin{equation}
    V \sim \frac{(T+1)\exp\left(-\frac{1+T}{T}\hat{X}\right)}{1+T\exp\left(-\frac{1+T}{T}\hat{X}\right)}.
    \label{eq: velocity profile}
\end{equation}

\subsection{Displacement}

Our results can also be used to derive a two-term asymptotic expansion for the total displacement of the free end. We begin by rewriting the exact mass balance Eq.\,\eqref{eq: mass balance} as  
\begin{equation}
X_c = \int_{0}^{\mathcal{L} - X_c} \left(H - 1\right)\, d\hat{X}.
\label{eq: mass balance long}
\end{equation}
Since our results \eqref{eq: sol_asymptotic} indicate that $H-1$ decays exponentially fast, we expect the integral to remain bounded as we take $\mathcal{L}\to\infty$. Accordingly, for $\mathcal{L}=\infty$, we have
\begin{equation}
X_c = \int_{0}^{\infty} \left(H - 1\right)\, d\hat{X}.
\end{equation}

Finally, evaluating the integrals using (\ref{eq: sol_asymptotic}a--b) and \eqref{eq: vel expressions}, we derive the two-term approximation
\begin{equation}
X_c \sim T + 1.
\label{eq: Xc approx}
\end{equation}



\section{Long finite sheets $(\mathcal{L}\gg 1, \, T\gg1)$}
\label{sec: phase changes}

The late-time analysis in the previous section is based on assuming a film that is semi-infinite for all $T$, 
\textit{i.e.} $\mathcal{L} = \infty$. For large but finite $\mathcal{L}$, the length of the sheet is $\mathcal{L} - X_c$, which decreases over time. Accordingly, the analysis in Sec.\,\ref{sec: long sheet} is  valid for $\mathcal{L} - X_c \gg 1$. Moreover, since $X_c \sim T$ for $T \gg 1$ (see Eq.\,\eqref{eq: Xc approx}), this condition is equivalent to $\mathcal{L} \gg T \gg 1$.

When $T$ becomes comparable to $\mathcal{L}$, the finite size of the sheet begins to influence the dynamics. Indeed, our numerical results for $\mathcal{L} = 20$ in Fig.\,\ref{fig: u_time}(c) show that when $T \approx 20$, the retraction speed decreases abruptly from unity to the value predicted by Eq.\,\eqref{eq: u asymp short}.  This marks the crossover from the semi-infinite sheet regime in Sec.\,\ref{sec: long sheet} to the finite-sheet late-time regime in Sec.\,\ref{sec: late phase}.

\begin{figure}[!t]
\begin{subfigure}{0.45\textwidth}
\centering
\includegraphics[width=\textwidth]{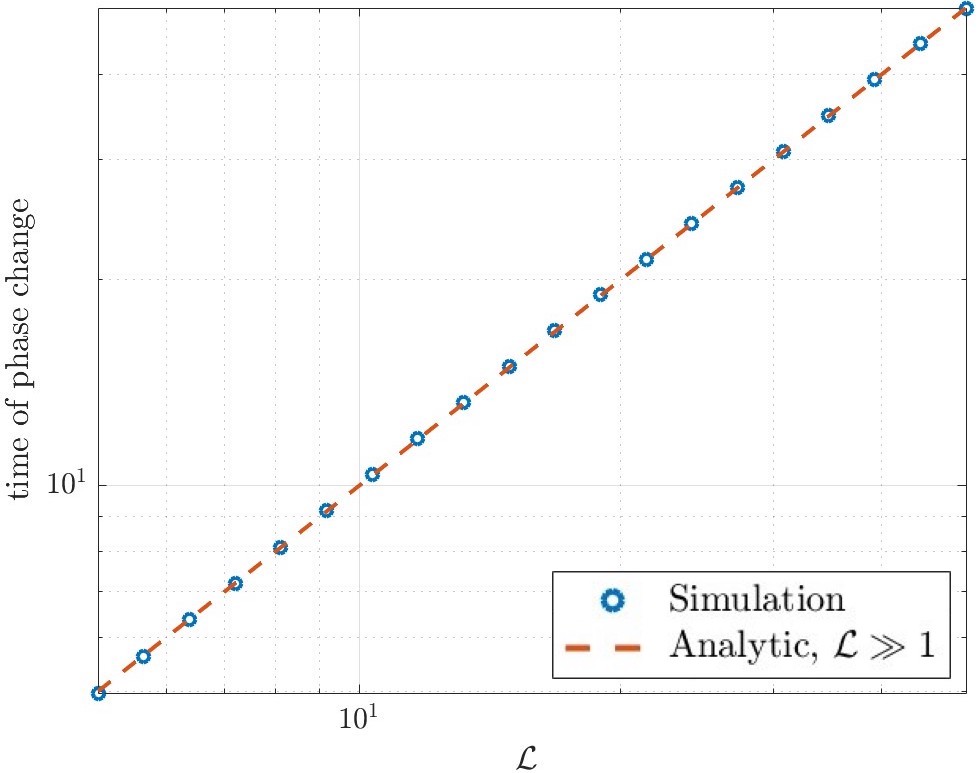}
\caption{$T_c$ vs.\ $\mathcal{L}$}
\end{subfigure}
~
\begin{subfigure}{0.45\textwidth}
\centering
\includegraphics[width=\textwidth]{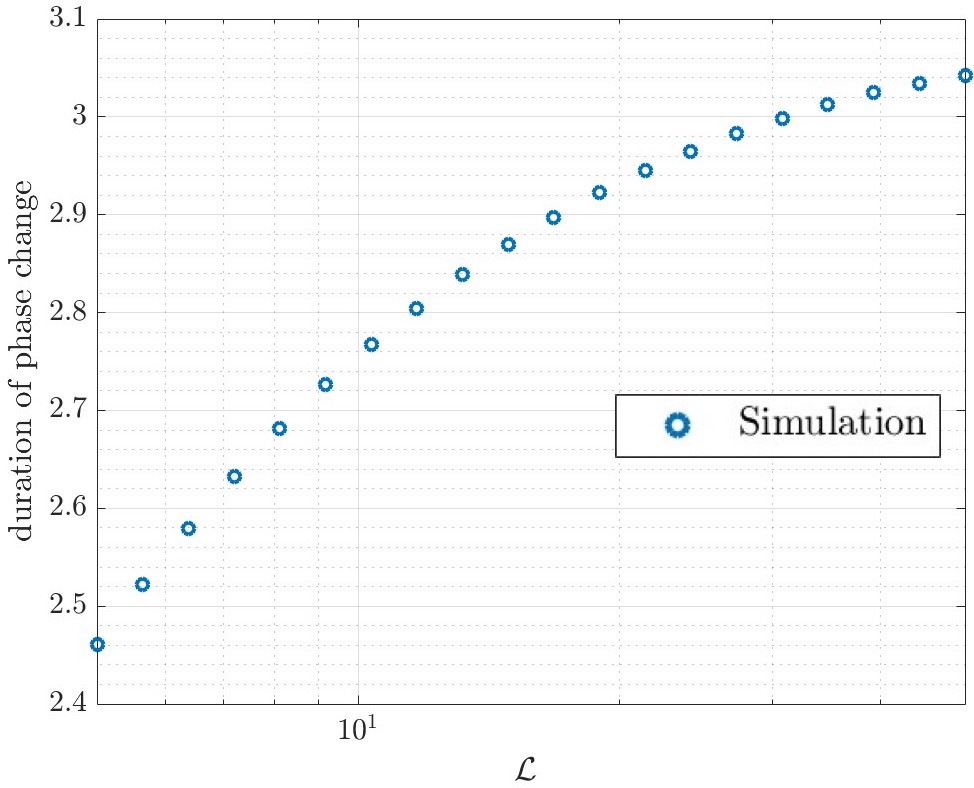}
\caption{$\Delta T_c$ vs.\ $\mathcal{L}$}
\end{subfigure}
\caption{
For large $\mathcal{L}$, the sheet undergoes a rapid deceleration during retraction, crossing over from the Taylor--Culick speed to the regime described by Eq.\,\eqref{eq: u asymp short}. (a) Crossover time $T_c$ when this deceleration occurs, and (b) duration $\Delta T_c$ of this transition,  shown for various values of $\mathcal{L}$. The dashed red line represents $T_c = \mathcal{L}$. 
}
\label{fig: t_duration}
\end{figure}

We numerically quantify this transitory regime by identifying the time $T_c$ at which the retraction speed $U(T)$ falls most sharply. In our simulations, this is given by the minimum of the derivative $dU/dT$. The duration of the transition,  $\Delta T_c$, is estimated as the full width at half maximum (FWHM) of this minimum. The resulting values of $T_c$ and $\Delta T_c$ as functions of $ \mathcal{L}$ are plotted in Fig.\,\ref{fig: t_duration}.

A simple estimate for the crossover time $T_c$ can be found by equating the leading-order approximation of $X_c$ in Eq.\,\eqref{eq: Xc approx} to $\mathcal{L}$, giving $T = \mathcal{L}$. This approximation is depicted by the dashed red line in Fig.\,\ref{fig: t_duration}(a). To further analyze the crossover regime, one could introduce a stretched timescale $T = \mathcal{L} + \tau(\mathcal{L}) T'$, where $\tau(\mathcal{L})$ is the characteristic timescale of the crossover. While we shall not pursue this direction further, we note that our numerics for $\Delta T_c$ in Fig.\,\ref{fig: t_duration}(b) suggest that $\tau(\mathcal{L})$ depends only weakly on $\mathcal{L}$.

\section{Concluding remarks}

We have analyzed the capillary-driven retraction of a finite, planar liquid sheet of initial length $l_0$ and thickness $h_0$, focusing on the regime where the sheet is initially thin $l_0/h_0\gg 1$ and highly viscous $\text{Oh}\gg 1$.
In this asymptotic regime, the fluid domain decomposes into two distinct regions: a small tip region at the free edge and a thin-film region that is assumed to be much longer than the tip. In the tip region, inertia is negligible, and the flow satisfies a similarity solution of the Stokes equations, which has been previously derived by Billingham \cite{Billingham:05} and Munro and Lister \cite{Munro:18}. In the thin-film region, the flow satisfies the standard one-dimensional mass and momentum equations for thin viscous films \cite{Erneux:93, Howell:96}, and is characterized by a single dimensionless parameter $\mathcal{L}= l_0/4 h_0 \text{Oh}$.

Asymptotic matching between the tip and thin-film regions yields an effective boundary condition at the free edge of the sheet, which physically represents a force balance between viscous stress and surface tension. We find that, although surface tension drives the flow, it enters the thin-film problem only via this boundary condition, with capillary stresses arising from thickness variations in the thin-film region being subdominant relative to viscous stresses. Consequently, only viscous and inertial terms appear at leading order in the thin-film momentum equation.

The thin-film problem was further simplified by identifying a locally conserved quantity that relates the horizontal velocity to the sheet thickness. This allowed us to reduce the coupled mass and momentum equations to a single equation governing either the film thickness or the velocity. In terms of thickness, the problem became mathematically equivalent to a Stefan-type problem, with the thickness satisfying the heat equation with a moving boundary at the free edge and time-dependent boundary conditions. Alternatively, in terms of velocity, the problem reduced to the Burgers’ equation, again with a moving boundary and time-dependent boundary conditions.

We numerically solved the reduced problem for the sheet thickness, revealing distinct dynamics that depend on the dimensionless parameter $\mathcal{L}$ and the rescaled time $T$, The results show a good agreement with previous studies of highly viscous liquid sheets \cite{Brenner:99, Savva:09}, including full Navier–Stokes simulations by Deka and Pierson \cite{Deka:20}.

To further illuminate our numerical results, we analyzed the reduced thin-film model in different asymptotic regimes, obtaining analytical approximations for the thickness and velocity profiles, as well as the retraction speed. For finite $\mathcal{L}$, the retraction speed scales as $2\sqrt{T/\pi}$ at early times and as $\mathcal{L}/(1 + T)^2$ at late times. These early- and late-time results agree with those of Savva and Bush \cite{Savva:09} and Deka and Pierson \cite{Deka:20}, respectively, though their derivations were based on different governing equations.

We have also derived closed-form asymptotic formulae in the intermediate regime $\mathcal{L} \gg T \gg 1$, which arises in our simulations for sufficiently long sheets. Interestingly, the retraction speed approaches the Taylor–Culick value given in Eq.\,\eqref{eq: taylor culick speed}, despite the sheet geometry being very different from that considered by Taylor \cite{Taylor:59} and Culick \cite{Culick:60}. In those classical studies, the sheet was assumed to develop a large circular rim, and the retraction speed was obtained by applying mass conservation together with a force balance between inertia and surface tension at the rim. In the present case, however, there are no assumptions on the vicinity of the free edge and the retraction speed instead emerges from the dynamics in the thin-film region far from the tip.

Our asymptotic results also allow us to estimate the regime of validity of our model. As time increases, the sheet simultaneously thickens and shortens, eventually invalidating the scale separation between the tip and thin-film regions that underpins our analysis. In detail, the similarity solution in Appendix~\ref{app: tip region} shows that the characteristic length scale of the tip region grows linearly with time, scaling as $\gamma t / 4\mu$ in dimensional variables. In contrast, the late-time analysis in Sec.\,\ref{sec: late phase} reveals that the horizontal extent of the thin-film region decreases as $\mathcal{L}/T$ for $T \gg 1$, or equivalently as $4\mu h_0 l_0 / \gamma t$ in dimensional variables.
These two lengthscales become comparable when $t = \mathcal{O}(\mu \sqrt{h_0 l_0} / \gamma)$, or $T = \mathcal{O}(\sqrt{l_0/h_0})$.  At this point, the tip and thin-film regions effectively merge into a single region. A scaling argument shows that inertia remains negligible in this regime, and the flow is governed by the full two-dimensional Stokes equations in a finite domain.

If the sheet is sufficiently long for the regime described in Sec.\,\ref{sec: long sheet} to apply, a different transition may occur at later times. For very long sheets, we found that the appropriate 
lengthscale of the thin-film flow is $l_I = 4 h_0 \text{Oh}$ (see Eq.\,\eqref{eq: li and ti}a). It becomes comparable to the tip length scale when $t = \mathcal{O}(\mu \text{Oh} / \gamma)$, or equivalently $T = \mathcal{O}(\text{Oh})$. As shown in Appendix~\ref{app: tip region}, this timescale also marks the onset of inertial effects in the tip region. When this regime is reached, the tip and thin-film regions again merge into a single region, but the flow is now governed by the full two-dimensional Navier–Stokes equations in an effectively infinite sheet---we refer to this as the two-dimensional inertial regime. For sufficiently long sheets, we expect the dynamics to eventually be described by the classical Taylor–Culick picture \cite{Taylor:59, Culick:60}, with most of the fluid accumulating in a large circular rim.

In summary, the asymptotic model based on the thin-film approximation  breaks down and the flow transitions into one of two regimes, depending on the relative magnitudes of $\sqrt{l_0/h_0}$ and $\text{Oh}$. If the sheet is not too long (or  alternatively,  sufficiently viscous), such that $\sqrt{l_0/h_0} \ll \text{Oh}$, the transition occurs at $T = \mathcal{O}(\sqrt{l_0/h_0})$, leading to the Stokes regime. If the sheet is sufficiently long (or alternatively, not sufficiently viscous), such that $\sqrt{l_0/h_0} \gg \text{Oh}$, the breakdown occurs instead at $T = \mathcal{O}(\text{Oh})$, leading to the inertial regime. Extending the present theory to either regime is challenging, as the dominant features of the flow are no longer captured by thin-film equations and require solving the full two-dimensional flow.

A more direct extension of our work would be to consider ruptured films with different geometries, where we anticipate that the fluid domain can be decomposed into tip and thin-film regions, provided the film is sufficiently slender and viscous. Examples include the expansion of a hole in a film and the retraction of axisymmetric liquid filaments, both of which have attracted significant experimental and theoretical interest \cite{Debregeas:95, Debregeas:98, Savva:09, Planchette:19, Pierson:20}. Another generalization would be to study films with initially non-uniform thickness profiles, such as linear or parabolic \cite{Keller:83}. The latter case was previously investigated by Munro \textit{et al.} \cite{Munro:15}, who also employed an asymptotic decomposition of the fluid domain---analogous to the one adopted here---in their analysis. In these generalized settings, an interesting open question is whether the thin-film flow still possesses a conserved quantity, as in the present case.

\appendix

\section{The tip region}
\label{app: tip region}

Here, we analyze the tip region of the retracting sheet. In this region, the slope of the fluid interface is of order unity, so its longitudinal extent is comparable to the sheet thickness $h_0$. Hence, the velocity scales as $h_0/t^*$ and the Reynolds number is of order $\frac{\rho h_0^2}{\mu t^*} = \frac{1}{4 \text{Oh}}\ll1$, so that inertia can be neglected. However, as we shall see, the tip region grows over time, so for inertial effects to remain negligible, we must also impose a condition on time. This is discussed further at the end of this Appendix section.

To analyze the flow, we adopt the same notation as in Sec.\,\ref{sec: statement}. We employ a reference frame moving with the tip and introduce the dimensionless coordinates 
\begin{equation}
\left(\overline{X},\overline{Y}\right) = \left(\frac{x-x_c(t)}{h_0}, \frac{y}{h_0}\right),
\end{equation}
and the unit vectors $\be_x$ and $\be_y$ in the $\overline{X}$ and $\overline{Y}$ directions, respectively. Note that, in the asymptotic limit $\mathcal{L} \gg 1$, the geometry of the tip region is semi-infinite, with $0 \leq \overline{X} < \infty$ and $|\overline{Y}| \leq \overline{H}$.
We normalize time, thickness profile, and velocity field
as in Eq.\,\eqref{eq: normalization},
and the stress tensor is normalized by $\gamma/h_0$. The dimensionless thickness profile, velocity field and stress tensors in the tip region are, respectively, denoted by $\overline{H}$, $\overline{\mathbf{V}}$, and $\overline{\boldsymbol{\Sigma}}$. The flow satisfies the Stokes equations
\begin{equation}
 \bnabla \bcdot \overline{\mathbf{V}} = 0, \qquad \bnabla \bcdot \overline{\boldsymbol{\Sigma}} = \boldsymbol{0},
 \label{eq: Stokes}
\end{equation}
subject to the kinematic boundary condition
\begin{equation}
\overline{\mathbf{V}}\cdot\bn    =  \left[1 + \left(\frac{\partial \overline{H}}{\partial \overline{X}}\right)^2\right]^{-1/2} \frac{\partial \overline{H}}{\partial T}  \quad \text{at}\quad |\overline{Y}|= \overline{H},
\label{eq: bc kin 2}
\end{equation}
and the dynamic boundary condition
\begin{equation}
\bn\bcdot\overline{\boldsymbol{\Sigma}} = \mathcal{K}\bn \quad \text{at}\quad |\overline{Y}|= \overline{H},
\label{eq: bc dyn 2}
\end{equation}
wherein $\mathcal{K}$ and $\bn$ are, respectively, the curvature and the outward normal of the fluid interface. 

Leading-order matching of the thickness profile in the tip and thin-film regions requires
\begin{equation}
\label{eq: thickness match}
\underbrace{\lim_{\overline{X}\to\infty} \overline{H}(\overline{X},T)}_{\text{tip}} = \underbrace{\lim_{X\to X_c} H(X) = H(X_c)}_{\text{thin film}},
\end{equation}
which is equivalent to applying Prandtl's matching rule between the two regions \cite{Hinch:book}. Similarly, leading-order matching  of the velocity field gives 
\begin{equation}
\underbrace{\lim_{\overline{X}\to\infty}\overline{\mathbf{V}}(\overline{X},\overline{Y},T)}_{\text{tip}} = \underbrace{V(X_c, T)\be_x}_{\text{thin film}},
\end{equation}
and leading-order matching of the stress field gives
\begin{equation}
\label{eq: stress match}
\underbrace{\lim_{\overline{X}\to\infty}\overline{\mathbf{\Sigma}}(\overline{X},\overline{Y},T)}_{\text{tip}} = \underbrace{\frac{\partial V}{\partial X} (X_c,T)\be_x\be_x}_{\text{thin film}},
\end{equation}
The evaluation of the velocity and stress fields in the thin-film region follows the standard approach outlined in Sec.\,\ref{sec: statement}, with further details given in \cite{Erneux:93,Howell:96}.

\subsection{Force balance and effective boundary condition}

We now examine the \textit{global force balance} in the tip region. To this end, we integrate Eq.\,\eqref{eq: Stokes} over the fluid domain with $0 \leq \overline{X} \leq \Lambda,$ and $|\overline{Y}| \leq \overline{H}(\Lambda)$, where the constant $\Lambda>0$ is arbitrary. Then, applying the divergence theorem yields
\begin{equation}
\int_{-\overline{H}(\Lambda)}^{\overline{H}(\Lambda)} \be_x\cdot \overline{\boldsymbol{\Sigma}} \, d\overline{Y} 
+ \int_{\substack{|\overline{Y}| = \bar{H}(\bar{X}) \\ 0 \leq \bar{X} \leq \Lambda }} \bn \cdot \overline{\boldsymbol{\Sigma}} \, dS 
= \boldsymbol{0},
\end{equation}
where $dS$ and $\bn$ are, respectively, the line element and outward normal of the fluid interface. Substituting Eq.\,\eqref{eq: bc dyn 2} into the second integral above and using the Frenet-Serret equations, we find
\begin{equation}
\int_{-\overline{H}(\Lambda)}^{\overline{H}(\Lambda)} \be_x\cdot \overline{\boldsymbol{\Sigma}} \, d\overline{Y} 
 + 2 \mathbf{\bt} = {\bf 0},
\label{eq: force balance}
\end{equation}
where $\mathbf{\bt}$ is the unit tangent vector of the interface at position $\overline{X}=\Lambda$.

We now take the limit $\Lambda \to \infty$ in Eq.\,\eqref{eq: force balance}.  
Using the matching conditions in Eqs.\,\eqref{eq: thickness match} and \eqref{eq: stress match}, the boundary integral simplifies to  
\begin{equation}
\int_{-\overline{H}(\Lambda)}^{\overline{H}(\Lambda)} 
\be_x \cdot \overline{\boldsymbol{\Sigma}} \, d\overline{Y}  
\;\longrightarrow\; 
2H(X_c,T)\,\frac{\partial V}{\partial X}(X_c,T)\be_x
\quad \text{as} \quad \Lambda \to \infty,
\label{eq: force balance 2}
\end{equation}
thereby expressing the force balance in the tip region in terms of the thickness and stress in the thin-film region. Since the thickness approaches a uniform value in this limit, the interfacial tangent vector satisfies $\bt \to \be_x$.  Taking the projection the resulting equation onto the $\mathbf{e}_x$ direction and simplifying furnishes the relation
\begin{equation}
H\,\frac{\partial V}{\partial X} = -1 
\quad \text{at} \quad X = X_c,
\end{equation}
which corresponds to the effective boundary condition given in Eq.\,\eqref{eq: bc free dim} (or, in dimensional form, Eq.\,\eqref{eq: dimens dyn}).

\subsection{Self-similar solution}

The Stokes flow problem in the tip region admits a self-similar solution, which has been numerically calculated by Billingham \cite{Billingham:05} and Munro and Lister \cite{Munro:18} (a three-dimensional, axisymmetric version of the problem, which also admits a self-similar solution, has been studied by Eggers \cite{Eggers:14}). To write the problem in self-similar form, we first note that Eq.\,\eqref{eq: alternative h cond} implies that the value of the far-field thickness is $H(X_c,T)= 1 + T$. With that motivation, we introduce the self-similar coordinates $(\chi,\eta) = (\overline{X},\overline{Y})/(1 + T)$ and thickness profile $\mathcal{H}\left(\chi,\eta\right)=\overline{H}/(1 + T)$. The Stokes equations Eq.\,\eqref{eq: Stokes} and dynamic boundary condition Eq.\,\eqref{eq: bc dyn 2} remain unaltered by this transformation, and the dynamic condition Eq.\,\eqref{eq: bc kin 2} transforms to
\begin{equation}
\overline{\mathbf{V}}\cdot \bn   =  \left[1 + \left(\frac{\partial \mathcal{H}}{\partial \chi}\right)^2\right]^{-1/2} \left(\mathcal{H} - \chi\frac{\partial \mathcal{H} }{\partial \chi} \right) \quad \text{at}\quad \eta= \mathcal{H} .
\end{equation}
The details of the self-similar solution can be found in \cite{Billingham:05} and \cite{Munro:18}. 

\subsection{Inertial effects}

The self-similar solution reveals that, as $T$ increases, the characteristic lengthscale of the tip region grows as $h_0 T = h_0 t/t^*$ while the characteristic velocity remains $h_0/t^*$. The effective Reynolds number therefore is $\frac{\rho h_0^2 t}{\mu t^{*2}} = \frac{T} {4 \text{Oh}}$. Hence, inertial effects remain negligible in the tip region as long as $T\ll \text{Oh}$.

\section{Reduced problem for the velocity profile $V$}
\label{app: v problem}

For completeness, we derive a reduced problem involving only the velocity profile $V$. To this end, we rewrite Eq.\,\eqref{eq: momentum dim} as
\begin{equation}
\frac{\partial V}{\partial T} + V\frac{\partial V}{\partial X} 
= \frac{\partial^2 V}{\partial X^2} + \frac{1}{H} \frac{\partial H}{\partial X} \frac{\partial V}{\partial X}.
\end{equation}
Substituting Eq.\,\eqref{eq: h and v rel} into the second term on the right-hand side and rearranging the terms, we arrive at
\begin{equation}
\frac{\partial V}{\partial T} + 2V\frac{\partial V}{\partial X} 
= \frac{\partial^2 V}{\partial X^2},
\end{equation}
which is the Burgers' equation \cite{Ockendon:03}. The velocity profile is subject to the initial conditions (\ref{eq: initial dim}b) and (\ref{eq: initial dim}c), the impermeability condition (\ref{eq: bc fixed}b), and the free-end boundary conditions Eq.\,\eqref{eq: bc free dim}, which, upon substitution of Eq.\,\eqref{eq: h and v rel}, become
\refstepcounter{equation}
$$
\frac{\partial V}{\partial X}(X_c(T), T) = -\frac{1}{1+T}, \qquad V(\mathcal{L}, T) = 0.
\eqno{(\theequation\mathrm{a},\mathrm{b})}
$$
Here, the location of the free edge, $X_c(T)$, evolves according to
\begin{equation}
\frac{dX_c}{dT} = V(X_c(T), T).
\end{equation}

Note that it is not necessary to solve both reduced problems for the thickness and velocity profiles, as the two fields are trivially related by Eq.\,\eqref{eq: h and v rel}. In the main text, we have opted to solve only the problem for $H$, since the governing equations are linear and thus more amenable to analysis.

\section{Conserved quantity in Lagrangian variables}\label{app: lagrangian}

We derive here the conservation law, Eq.\,\eqref{eq: conserved}, using a Lagrangian formulation of the flow. Consider an infinitesimal fluid element that initially has thickness $h_0$ and length $d\xi$ in dimensional form.
At time $t$, the element deforms, changing its thickness to $h$
and its length to $dx$. Mass conservation then requires $h\, dx = h_0\, d \xi$.

Using $\xi$ as the Lagrangian coordinate, Eqs.\,\eqref{eq: continuity} and \eqref{eq: dimen mom} transform into
\refstepcounter{equation}\label{eq: lag equation}
$$
\pd{h}{\xi} = - \frac{h^2}{h_0}\pd{v}{\xi}, \qquad \rho h \pd{v}{t} = 4 \mu \frac{h}{h_0}\pd{}{\xi}\left(\frac{h^2}{h_0}\pd{v}{\xi}\right),
\eqno{(\theequation\mathrm{a},\mathrm{b})}
$$
where capillary stresses have been neglected, as discussed in Sec.\,\ref{sec: scales}.  Note that in the Lagrangian description, material derivatives reduce to partial derivatives.

From Eq.\,(\ref{eq: lag equation}a), the viscous stress is related to the rate of thinning by $4\mu \frac{h^2}{h_0}\pd{v}{\xi}=-4\mu \pd{h}{t}$. Substituting this into Eq.\,(\ref{eq: lag equation}b) gives
\begin{equation}
\rho h \pd{v}{t} = -4 \mu \frac{h}{h_0}\pd{}{\xi}\left(\pd{h}{t}\right) \Rightarrow \pd{v}{t} = -\frac{4 \mu }{\rho h_0}\pd{}{\xi}\left(\pd{h}{t}\right).
\end{equation}
The left-hand side is precisely the Lagrangian acceleration, and the right-hand side corresponds to the viscous force per unit mass of the fluid element. Exchanging the order of time and spatial derivatives and combining terms, we obtain the conservation law
\begin{equation}
\pd{}{t}\left(v +\frac{4 \mu }{\rho h_0}\pd{h}{\xi}\right) = 0,
\end{equation}
which is precisely Eq.\,\eqref{eq: conserved} when written in terms of the original dimensionless variables.

\section{Numerical Method}
\label{app: numerics}
\subsection{Conversion to Static Frame Equations}
To handle the time-varying domain and boundary conditions, we transformed the original problem by introducing the rescaled variables
\refstepcounter{equation}
$$
X' = \frac{\mathcal{L} - X}{\mathcal{L} - X_c(T)},\quad H'(X',T) = \frac{H(X,T)}{1+T},\quad S = (\mathcal{L} - X_c(T))^2
\eqno{(\theequation\mathrm{a},\mathrm{b}, \mathrm{c})}.
$$
which fixes the horizontal coordinate to $0<X'<1$ and keeps the thickness at $X'=0$ bounded in time.

The resulting dimensionless formulation, suitable for numerical implementation, is given by the partial differential equation,\begin{equation}
\frac{\partial H'}{\partial T} = \frac{1}{S(T)}\frac{\partial^2 H'}{\partial X'^2}+ \frac{X'}{2S(T)}\frac{dS(T)}{dT} \frac{\partial H'}{\partial X'}- \frac{H'}{T+1};
\label{eq: numerics-h}
\end{equation}
the boundary conditions
\refstepcounter{equation}
$$
\frac{\partial H'}{\partial X'}(0, T) = 0,\quad H'(1, T) = 1;
\eqno{(\theequation\mathrm{a},\mathrm{b})}.
\label{eq: numerics-bc}
$$
and the initial condition
\begin{equation}
H'(X', 0) = 1.
\end{equation}
The evolution equation for the moving boundary variable $S(T)$,
\begin{equation}
\frac{dS}{dT} = -2\frac{\partial H'}{\partial X'}(1, T), \quad S(0)=\mathcal{L}^2;
\label{eq: numerics-s}
\end{equation}

\subsection{Description of the Algorithm}
Our simulations typically employed $1000$ grid points on the solution domain $X' \in [0,1]$, corresponding to a typical grid size of order $\Delta X' \sim 10^{-3}$. 
The time step was generally of order $\Delta T \sim 10^{-4}$. 
Spatial derivatives of $H'(X',T)$ were computed using central difference approximations. At any given time step, the full state of the system is represented by the pair $(H',S)$.  Time-stepping this pair from $T$ to $T+\Delta T$ was carried out using a modified Crank--Nicolson scheme, developed to address the unique features of our problem. We describe these modifications below. 

First, updating $S(T)$ requires evaluating $\partial_{X'} H'$ at the boundary point $X'=1$. However, since there are no grid points beyond $X'=1$, a direct application of a central difference approximation is not possible, which would otherwise yield second-order accuracy in $\Delta X'$. To address this, we express $\partial^2 H'/\partial X'^2 $ at $X'=1$ in terms of $S(T)$ by combining Eqs.\,\eqref{eq: numerics-bc} and \eqref{eq: numerics-h}. This yields
\begin{gather}
    \frac{\partial^2 H'}{\partial X'^2}(X'=1,T) 
    = \frac{1}{4} \left(\frac{dS}{dT}\right)^2 + \frac{S}{1+T}.
    \label{eq: s-issue-1}
\end{gather}
Next, we perform a Taylor series expansion for $\partial_{X'} H'$ near $X'=1$:
\begin{gather}
    \frac{dS}{dT} = 
    -2\,\frac{\partial H'}{\partial X'}\left(1-\frac{\Delta X'}{2},T\right) 
    - \Delta X'\,\frac{\partial^2 H'}{\partial X'^2}(1,T),
    \label{eq: s-issue-2}
\end{gather}
where both $X'=1$ and $X'=1-\Delta X'$ are grid points. 
Thus, the term $\partial H'/\partial{X'}(1-\Delta X'/2, T)$ can be computed to second-order accuracy using a central difference approximation.

For the Crank--Nicolson update of $S(T)$, we require the value of $dS/dT$ at the mid-point in time, $T+\Delta T/2$:
\begin{gather}
    S(T+\Delta T) = S(T) + \Delta T\,\frac{dS}{dT}\left(T+\frac{\Delta T}{2}\right) + \mathcal{O}(\Delta T^2).
\end{gather}

Substituting Eqs.\,\eqref{eq: s-issue-1} and \eqref{eq: s-issue-2}, and performing an order-of-magnitude analysis, we obtain
\begin{gather}
    \frac{dS}{dT}\left(T+\frac{\Delta T}{2}\right)
    = -2\mathfrak{h}-\left(\mathfrak{h}^2+\frac{S(T)}{1+T}\right)\Delta X',
\end{gather}
where
\begin{gather}
    \mathfrak{h}:= \frac{\partial H'}{\partial X'} =
    \frac{1 - \tfrac{1}{2}H'(1-\Delta X',T) - \tfrac{1}{2}H'(1-\Delta X',T+\Delta T)}{\Delta X'}.
\end{gather}
This formulation is accurate to second order in both $\Delta X'$ and $\Delta T$.

Another complication arises from the Crank--Nicolson scheme itself. 
The update rules for $H'$ and $S$ can be formally written as
\begin{equation}
H'_f = \mathrm{H'\_step}(H'_n, S_n, S_f),
\quad
S_f = \mathrm{S\_step}(S_n, H'_n, H'_f),
\end{equation}
where the subscript $n$ denotes the present state and $f$ denotes the future state.
However, during a time step, neither $H'_f$ nor $S_f$ is initially known, which creates a coupling problem. To resolve this, we employed the following iterative procedure:
\begin{center}
\begin{enumerate}
    \item $S_n \;\rightarrow\; S_f = \mathrm{S\_step}(S_n, H'_n, H'_n)$
    \item $H'_n \;\rightarrow\; H'_f = \mathrm{H'\_step}(H'_n, S_n, S_f)$
    \item $S_n \;\rightarrow\; S_f = \mathrm{S\_step}(S_n, H'_n, H'_f)$
    \item $H'_n \;\rightarrow\; H'_f = \mathrm{H'\_step}(H'_n, S_n, S_f)$
\end{enumerate}
\end{center}
In Step 1, the current value $H'_n$ is temporarily used in place of $H'_f$ to compute a first-order accurate estimate of $S_f$.  This estimate is then used in Step 2 to obtain a more accurate $H'_f$.  Next, this updated $H'_f$ is used in Step 3 to compute a second-order accurate $S_f$, which is then employed in Step 4 to obtain a second-order accurate $H'_f$. This iterative procedure ensures second-order accuracy for both $H'$ and $S$ in time.

We share our simulation codes on GitHub~\cite{GitHub}.

\section{Late-time WKB approximation for infinitely-long sheets ($\mathcal{L}=\infty$, $T\gg1$)}
\label{app: velocity alternative}

In Sec.\,\ref{sec: long sheet}, we employed a regular expansion in powers of $1/T$ to calculate the thickness profile for $\mathcal{L} = \infty$ and $T \gg 1$. We found that the resulting expansion, Eq.\,\eqref{eq: expansion nonuni}, is non-uniform in $\hat{X}$. Specifically, the exponential terms in Eq.\,\eqref{eq: expansion nonuni} become comparable when $X - X_c = \hat{X} = \mathcal{O}(T)$, signaling the breakdown of the asymptotic series in that region.

Here, we employ the WKB method \cite{Hinch:book} to derive an approximation that remains valid for $\hat{X} = \mathcal{O}(T)$. To this end, we define the rescaled coordinate
\begin{equation}
\breve{X} = \left(X - X_c\right)/T,
\label{eq: wkb coordinate}
\end{equation}
and write the thickness profile using the WKB \textit{ansatz},
\begin{equation}
H(X,T) = 1 + T\exp\left(-T\Phi(\breve{X},T)\right),
\label{eq:H ansatz}
\end{equation}
where we have introduced the auxiliary function $\Phi(\breve{X},T)$.
Substituting Eqs.\,\eqref{eq: wkb coordinate} and \eqref{eq:H ansatz} into Eq.\,\eqref{eq: diffusion h long} yields the (exact) equation
\begin{equation}
1  + \frac{\partial^2 \Phi}{\partial \breve{X}^2} +  T\left[ - \Phi+ \breve{X} \frac{\partial \Phi}{\partial \breve{X}}+ \frac{d X_c}{dT}\frac{\partial \Phi}{\partial \breve{X}}-\left( \frac{\partial \Phi}{\partial \breve{X}} \right)^2\right] - T^2 \frac{\partial \Phi}{\partial T}  = 0.
  \label{eq wkb}
\end{equation}
In terms of the auxiliary function, the boundary conditions \eqref{eq: alternative h cond} and \eqref{eq: long sheet bc 2} become
\begin{equation}
    \Phi = 0 \quad \text{at}\quad \breve{X}=0, \qquad \Phi \to \infty \quad \text{as}\quad \breve{X}\to\infty.
    \label{bc wkb}
\end{equation}

We now expand the retraction speed $dX_c/dT$ as in Eq.\,(\ref{eq: h_long_exp}a) using Eq.\,\eqref{eq: vel expressions}. To expand $\Phi$, we note that (\ref{eq: h_long_exp}a) shows that for $\breve{X}=\hat{X}/T\ll1$ we have that $T\Phi \sim   \hat{X} = T\breve{X}$, which shows that $\Phi$ must be of order unity. Accordingly,  we pose the regular expansion
\begin{equation}
\Phi(\breve{X},T) = \Phi_0\left(\breve{X}\right) + \frac{1}{T}\Phi_{-1}\left(\breve{X}\right) + \mathcal{O}\left(\frac{1}{T^2}\right).
\end{equation}

Expanding Eqs.\,\eqref{eq wkb} and  \eqref{bc wkb} to leading-order in $1/T$ yields
\begin{equation}
 - \Phi_0+ \breve{X} \frac{\partial \Phi_0}{\partial \breve{X}}+ \frac{\partial \Phi_0}{\partial \breve{X}}-\left( \frac{\partial \Phi_0}{\partial \breve{X}} \right)^2 = 0,
  \label{eq wkb lead}
\end{equation}
subject to 
\begin{equation}
    \Phi_0 = 0 \quad \text{at}\quad \breve{X}=0, \qquad \Phi_0 \to \infty \quad \text{as}\quad \breve{X}\to\infty.
    \label{bc wkb lead}
\end{equation}
We can verify that the appropriate solution is
\begin{equation}
    \Phi_0 = \breve{X}.
\end{equation}

Going to the next order, we find
\begin{equation}
1 - \Phi_{-1}+ \breve{X} \frac{\partial \Phi_{-1}}{\partial \breve{X}}- \frac{\partial \Phi_{-1}}{\partial \breve{X}} = 0,
  \label{eq wkb corr}
\end{equation}
subject to 
\begin{equation}
    \Phi_{-1} = 0 \quad \text{at}\quad \breve{X}=0, \qquad \Phi_{-1} \to \infty \quad \text{as}\quad \breve{X}\to\infty.
    \label{bc wkb corr}
\end{equation}
Once again, we can verify that the appropriate solution is 
\begin{equation}
    \Phi_{-1} = \breve{X}.
\end{equation}

Substituting the expansion into Eq.\,\eqref{eq:H ansatz}, we obtain the two-term WKB approximation for the thickness profile,
\begin{equation}
H(X,T) \sim 1 + T\exp\left(-\left(1 +T\right)\breve{X}\right),
\label{eq:H expansion}
\end{equation}
which is equivalent to Eq.\,\eqref{eq: wkb text} when written in terms of the shifted coordinate $\hat{X} = T\breve{X} = X - X_c$.

\bibliography{bibliography}

\end{document}